\documentclass[aps,pra,superscriptaddress,floatfix,showpacs,twocolumn]{revtex4-1}
\usepackage{amsmath}
\usepackage{graphicx}
\usepackage{mathtools}
\usepackage{amsfonts}
\usepackage{float}
\usepackage{comment}
\usepackage{xcolor}

\def\vv#1{{\textcolor{black}{{#1}}}}

\usepackage{natbib}
\usepackage{hyperref}
\usepackage{cleveref}
\newtheorem{theorem}{Theorem}

\newtheorem{assumption}{Assumption}
\newtheorem{definition}{Definition}
\newtheorem{cor}{Corollary}[theorem]

\newcommand{\Sym}[2]{\mathcal{S}^{#2}\left(\left\{{#1}\right\}\right)}
\newcommand{\SymPlus}[3]{\mathcal{S}^{#3}\left(\left\{#1\right\}#2\right)}

\newenvironment{proof}[1][Proof]{\noindent\textbf{#1.} }{\ \rule{0.5em}{0.5em}}
\newcommand{\del}{\ensuremath{\partial}}

\hypersetup{
	colorlinks,
	linkcolor={blue},
	citecolor={blue},
	urlcolor={blue!80!black}
}

\begin{document}
	\title{Provable bounds for the Korteweg-de Vries reduction in multi-component Nonlinear Schr\"odinger Equation}
	\author{Swetlana Swarup}
	\affiliation{Indian Institute of Science Education and Research, Kolkata, Mohanpur - 741246, India}
\affiliation{International centre for theoretical sciences, Tata Institute of Fundamental Research, Bangalore - 560089, India}
\affiliation{School of Physics and Astronomy,
University of Minnesota, Minneapolis, MN 55455, USA}
	\author{Vishal Vasan}
	\affiliation{International centre for theoretical sciences, Tata Institute of Fundamental Research, Bangalore - 560089, India}
	\author{Manas Kulkarni}
	\affiliation{International centre for theoretical sciences, Tata Institute of Fundamental Research, Bangalore - 560089, India}
	
	\begin{abstract}
		 We study the dynamics of multi-component Bose gas described by the Vector Nonlinear Schr\"{o}dinger Equation (VNLS), aka the Vector Gross--Pitaevskii Equation (VGPE) . Through a Madelung transformation, the VNLS can be reduced to coupled hydrodynamic equations in terms of multiple density and velocity fields. Using a multi-scaling and a perturbation method along with the Fredholm alternative, we reduce the problem to a Korteweg de-Vries (KdV) system. This is of great importance to study more transparently, the obscure features hidden in VNLS. This ensures that hydrodynamic effects such as dispersion and nonlinearity are captured at an equal footing. Importantly, before studying the KdV connection, we provide a rigorous analysis of the linear problem. We write down a set of theorems along with proofs and associated corollaries that shine light on the conditions of existence and nature of eigenvalues and eigenvectors of the linear problem. This rigorous analysis is paramount for understanding the nonlinear problem and the KdV connection. We provide strong evidence of agreement between VNLS systems and KdV equations by using soliton solutions as a platform for comparison. Our results  are expected to be relevant not only for cold atomic gases, but also for nonlinear optics and other branches where VNLS equations play a defining role. 
	\end{abstract}
	
\date{\today}

	\maketitle
	
	\section{\label{sec:I} Introduction}

Multi-component coupled systems are ubiquitous in physics ranging from cold atomic systems \cite{smerzi2003macroscopic, Burchianti2018dual,roati2007k,thalhammer2008double,ejnisman1998studies,wacker2015tunable,mccarron2011dual,
papp2008tunable,wang2015double,matthews1999watching} to nonlinear optics \cite{chen1997coupled,ostrovskaya1999interaction,mitschke1987experimental,hasegawa1980self,andrekson1991observation,mitchell1998observation,mitchell1997self}. Such systems are typically nonlinear, i.e., with considerable interactions and often have an intricate interplay between the various species. Given these rich interactions, both intra-species and inter-species, the cutting-edge technologies \cite{andrews1997propagation,andrews1996direct,anderson1995observation,mewes1996bose} to image their collective behaviour and the ability to engineer these systems makes them a rich platform to study far-from-equilibrium physics in multi-component systems. 

Often arriving at a Hamiltonian or a set of differential equations to describe the collective behaviour of particles is in itself a difficult task. However, a substantial work is done in this direction and there is a reasonable understanding of an effective Hamiltonian or differential equations that could describe multi-species systems in certain parameter regimes and conditions \cite{agrawal2000nonlinear,kevrekidis2008basic}. However, the complex nature of these systems results in dealing with equations which are often cryptic and the consequences of which are difficult to understand.  For example, if the collective behaviour of multi-species systems have nonlinearities and higher derivatives, one would expect to see nonlinear and dispersive effects. Such fingerprints of hydrodynamics are often completely elusive. Even the linearized version of the problem, existence of stable modes are unclear. Hence, it is of great importance to develop a systematic theory that will lead to a universal framework \cite{kulkarnipra,erdHos2007rigorous,dalfovo1999theory,zakharov1971korteweg,ablowitz2011nonlinear,pethick2008bose,horikis2014nls,leblond2008reductive,zakharov1986multi,spiegel1980fluid,rperturbation}, that captures the various hallmarks of collective field theory or hydrodynamics. 

We will describe here several systems where multi-component physics comes into play. Nonetheless, we will keep our main motivation as non-equilibrium dynamics in Bose mixtures \cite{Burchianti2018dual}. In the context of cold atoms, VNLS (aka multi-component Gross$-$Pitaevskii equation) appears in multiple coupled species of bosons \cite{liu2017criticality,zhou2008phase,liu2018vortex,kasamatsu2005modulation,kasamatsu2006modulation,mareeswaran2016superposed,kasamatsu2004multiple,pattinson2014phenomenological,feng2014general,manikandan2016manipulating,kuo2008phase,caliari2008spatial} or bosons with hyperfine degrees of freedom (spinor BECs \cite{wen2017three,sun2018matter,li2017non,liu2018vector,belobo2018exotic,massignan2015magnetism,oztas2019spin,yakimenko2009two,hong2011weak}). The nonlinearities or interactions one sees are of inter-species and intra-species type. Such systems can be placed out of equilibrium and their collective density dynamics can be imaged in-situ using cutting-edge technologies in absorption imaging techniques. On the other hand, non-linear optical setups also provide a great platform for studying multi-component NLS systems \cite{afanasyev1989dynamics,trillo1988optical}.  Here, typically, the role of time is played by an additional spatial axis \cite{agrawal2000nonlinear}. The intensity of light can be directly measured \cite{mitchell1996self,martienssen1964coherence} which is captured by a set of NLS equations. Apart from these two main avenues, NLS-type equations also appears in a variety of other contexts like Quantum Mechanics \cite{rosales1992non}, accelerator dynamics \cite{fedele1993thermal}, biomolecular dynamics \cite{davydov1985solitons,daniel2002soliton,qin2010solitonic}, plasma and water waves \cite{kourakis2006nonlinear,infeld2000nonlinear}. \\

Keeping in mind the above motivation, we start with the description of  $N$-component coupled NLS equations. Whenever appropriate, we will discuss the physical relevance mainly keeping collective description of cold atomic systems in mind. The $N$-component Nonlinear Schr\"{o}dinger Equation (NLS) in 1D is given by
	\begin{equation}
		i\hbar \frac{\del\psi_k}{\del t}=-\frac{\hbar^2}{2m}\frac{\del^2\psi_k}{\del x^2}+\sum_{j=1}^{N} \alpha_{kj}|\psi_j|^2 \psi_k
\label{mnls}
	\end{equation}
	where $\psi_k$ is the macroscopic wavefunction and $\alpha$ is the matrix of coupling constants. It is to be noted that the diagonal elements of the $\alpha$ matrix correspond to intra-species interaction and the off-diagonal elements correspond to inter-species interaction. {\vv{We assume a symmetric coupling and hence $\alpha$ is a symmetric matrix.} It turns out that in cold atomic systems, both intra-species and inter-species coupling are tunable via sweeping across a Feshbach resonance \cite{inouye1998observation,roati2007k,wacker2015tunable,thalhammer2008double}.  The quantity $|\psi_k|^2$ gives the density of particles of species type $k$ and the angle associated with the complex number $\psi$ gives the phase both of which are measurable in experiments. The equations have a Hamiltonian structure given by the Hamiltonian
	\begin{equation}
	\mathcal{H}=\int dx \sum_{k=1}^{N}\bigg(\frac{\hbar^2|\del_x \psi_k|^2}{2m}+\sum_{j=1}^{N} \frac{\alpha_{kj}}{2}|\psi_k|^2|\psi_j|^2 \bigg)
	\end{equation} 
	equipped with Poisson brackets $\{\psi_j^*(x),\psi_k(y)\}=\frac{i}{\hbar}\delta_{jk}\delta(x-y)$. The NLS equations \cite{ablowitzbookNLS} of motion can be obtained from 
	\begin{equation}
		\frac{\del \psi_k}{\del t}=\{\psi_k,\mathcal{H}\} \;\quad \forall \,k
	\end{equation}
	
It is to be pointed out that single component version has been studied intensively in literature both from a point of mathematical interest \cite{chiron2012travelling,spiegel1980fluid,huang2001dark,kamchatnov2012generation,yan2009exact,huang2001korteweg} and experimentally in physical systems \cite{anderson1995observation,andrews1997propagation,davis1995bose,andrews1996direct} and finds applications in a variety of fields \cite{rosales1992non,fedele1993thermal,davydov1985solitons,mollenauer1980experimental,chang2012extensive,kivshar1990dark,kuwamoto2004magnetic,kevrekidis2007emergent}. %
Typical aspects studied both theoretically and experimentally include non-equilibrium evolution of density profiles, solitons \cite{burger1999dark,khaykovich2002formation,kevrekidis2016solitons}, quenches \cite{franchini2016hydrodynamics}, problems in presence of defects and disorder \cite{paiva2015cooling,hulet2009tunable,chen2009experimental,chen2008phase}. In addition to systems which have a Hamiltonian structure, there has been a lot of work on  Driven-dissipative (gain-loss \cite{Smirnov:14,Kivshar:98,Sich:12,Amo:11,Menon:10,Hivet:12,Lagoudakis:08,Amo:09}) and PT-symmetric systems \cite{el2007theory,makris2008beam,ruter2010observation}. These are interesting open system or non-Hermitian generalizations of the NLS family of equations and are not a subject of our current work.
 

From \eqref{mnls} it is difficult to understand non-equilibrium phenomena. More precisely, given the initial condition $\psi_i(x,0)$, one is interested in the time evolution $\psi_i(x,t)$ from which various experimentally relevant quantities can be extracted. Experimentally, one can prepare an initial density profile $|\psi_i(x,0)|^2$ and associated phase (or its derivative which is akin to the velocity field). Then the system can be made to evolve and the time evolution of these quantities can be obtained. The central goal of our work is to provide a universal framework to understand the linear and non-linear properties of these evolving density and velocity fields of all species. A natural well-known conservative partial differential equation that is expected to capture fingerprints of hydrodynamics (namely, dispersion and nonlinearity) is the KdV equation. However, a systematic understanding of the linear problem and then the non-linear problem is far from obvious. Before going into the contents and details of our paper, below we provide a summary of our main findings.	

The main contribution of the present manuscript is the systematic derivation of the qualitative long time dynamics of equation \eqref{mnls}. In particular, we show that the evolution of small perturbations to the trivial state are governed by the KdV equation. {\vv{For the single component case ($N=1$), the derivation of KdV from NLS is well known  }\cite{gardner1965similarity,su1969korteweg,jeffrey1982asymptotic,johnson1997modern,newell1985solitons}}. The present work provides the systematic and explicit derivation of KdV for the multi-component case. The coefficients for the KdV and the speed of sound (\emph{i.e.} in the frame of reference for the evolution) are given in terms of the background trivial state, mass and the coupling coefficients $\alpha_{jk}$. We achieve this by a systematic and complete analysis of the spectral problem for the equation linearised about the trivial state. We derive necessary and sufficient conditions for real sound speeds dependent solely on the coupling matrix. For a specific case of the coupling matrix, we obtain necessary and sufficient conditions for sound speeds to be distinct. The coefficients of the KdV dynamics are given in terms of the eigenvectors of the linearised problem. For the case of repeated eigenvalues, we provide the eigenvectors explicitly. For the simple eigenvalue case, we construct the eigenvectors of the linearised problem in terms of the eigenvectors of the coupling matrix. We also provide an  accurate efficient and stable numerical algorithm to compute the eigenvectors of the coupling matrix individually that also provides us with information on how eigenvalues (sound speeds) and eigenvectors (KdV coefficients) change as the cross-component coupling coefficient varies.

Though the bulk of the present work analyses the multi-component system \eqref{mnls} under Assumption \ref{model_assumption}, our analysis is general enough to be readily extended to other coupling matrices. We have attempted to provide the reader with sufficient details of the proofs to make this generalisation obvious. We will consider alternative couplings in future works. We have also attempted to highlight the role of a key mathematical idea, namely the Fredholm alternative, in reductive perturbative theory. To this end, we present perhaps more detail than is typical so that the interested reader can appreciate the systematic nature of the derivation.

The contents of the paper are organized as follows. In \autoref{sec:IIA} the coupled NLS system \eqref{mnls} is transformed into hydrodynamic form where the relevant physical quantities are the density and velocity fields. The linear regime of the thus obtained hydrodynamic equations is studied in \autoref{sec:IIB}. Rescaling the independent variables, $x$ and $t$, so that we focus our attention on the dynamics on long space and time scales, we are lead to, at linear order, an equation that governs the linear stability of perturbations to a trivial background state. In \autoref{sec:IIC} we determine necessary and sufficient conditions for these perturbations to propagate stably in terms of the coupling matrix of the original system \eqref{mnls}. Mathematically, this involves a complete analysis of the eigen-system for a particular matrix $\mathcal A$. This analysis includes an explicit formula for the characteristic polynomial; necessary and sufficient conditions for repeated roots; and expressions for the associated eigenvectors. The eigenvalues and eigenvectors of $\mathcal A$ play an important role in the qualitative dynamics of perturbations to the trivial state:  (i) the eigenvalues determine the sound speed and (ii) eigenvectors determine coefficients of the effective KdV equation governing these perturbations. After introducing the relevant mathematical ideas in \autoref{sec:IID}, we proceed with the reductive perturbation method applied on the density and velocity fields to derive the effective KdV equation for $N$-components in \autoref{sec:III}. We  present explicit results for case of few components ($N=2,3$) in \autoref{sec:IV}. In \autoref{sec:V}, we discuss the results of numerical comparison between the coupled ($N=2$) NLS and corresponding KdV by simulating a solitary wave profile which further explicates nontrivial features of the NLS in the reduced KdV. We finally conclude in \autoref{sec:VI} along with an outlook. 

	\section{\label{sec:II} Hydrodynamic Model and Linearization}
	\subsection{\label{sec:IIA} Modeling}
	As mentioned in the introduction, one of the primary contributions of the present manuscript is a characterization of a multicomponent system, particularly in the small-amplitude long-wavelength regime. This regime naturally leads to a coupled multi-species KdV-like model. In order to derive the associated KdV model, we first perform the usual Madelung transform \cite{Madelung1927} to obtain a set of hydrodynamic equations for the density and velocity,
	\begin{equation}\label{eqn:madelung}
	\psi_k(x,t) =\sqrt{\rho_k(x,t)}e^{i(m/\hbar)\int_{0}^{x}v_k(x',t)dx'}
	\end{equation}
	where $\psi_k$ is the macroscopic wavefunction of the  $k^{th}$ condensate, $\rho_k(x,t)$ is the corresponding density field and $v_k(x,t)$, the velocity field where $k=1,\ldots N$. The resultant equations of motion are an equation of continuity (for the density) 
	\begin{equation}
		\frac{\del \rho_k}{\del t} + \frac{\del}{\del x} \big(\rho_k v_k\big)=0,
	\end{equation}
	and the Euler equation (for the velocity)
	\begin{equation}
	\frac{\del v_k}{\del t} = -\frac{\del}{\del x}\bigg[\frac{v_k^2}{2}+\frac{1}{m}\sum_{j=1}^{N}\alpha_{kj}\rho_j- \bigg(\frac{\hbar^2}{2m^2}\bigg)\frac{\del_x^2\sqrt{\rho_k}}{\sqrt{\rho_k}}\bigg].
	\end{equation}
	
	We remark that the density equation of $k-$th component is uncoupled; the coupling matrix only appearing in the velocity equations. It is worth emphasizing that the there is no approximation in obtaining the above equations. They are fully equivalent to the original multicomponent system.
	
	\subsection{\label{sec:IIB} Linearized dynamics}
	A trivial solution to the hydrodynamic equations is given by setting the densities to non-negative constants $\rho_{0k}$ and the velocities to zero. A natural question is how small perturbations to this background state evolve. The standard approach to this question is given by linearising the equations about the background state, namely setting $\rho_k=\rho_{0k}+\delta\rho_k$ and $v_k=\delta v_k$, to obtain the following linear evolution equations for the perturbations $\delta\rho_k,\delta v_k$
	\begin{subequations}\label{eqn:linear:hydro}
		\begin{gather}
			\frac{\del}{\del t}\delta\rho_k=-\rho_{0k}\frac{\del}{\del x}\delta v_k,\\
			\frac{\del}{\del t}\delta v_k=-\frac{1}{m}\frac{\del}{\del x}\sum_{j=1}^{N}\alpha_{kj}\delta\rho_j+\frac{\hbar^2}{4m^2\rho_{0k}}\frac{\del^3}{\del x^3}\delta\rho_k.
		\end{gather}
	\end{subequations}
	These constant coefficient equations are readily solved using Fourier methods. Furthermore, for the simpler case of $\alpha$ equal to a diagonal matrix (the uncoupled case), the above equations have solutions of the form $\mathrm{exp}(ipx+i\omega t)$ with $\omega\approx c_1p + c_2 p^3 $ for some real constants $c_1,c_2$. This indicates waves travel at speed $c_1$ but also disperse due to the presence of the cubic term. This suggests that if we consider long-wavelength perturbations such that the dispersive term balances with the nonlinear corrections, we may arrive at a KdV-like model. This same argument applies also to the coupled case, \emph{i.e.} for a generic matrix $\alpha$. 

	To obtain a balance between nonlinearity and dispersion, we assume the following form for the original full density and velocity: 
	\begin{align}\label{perturbation:rho}
	\rho_k &=\rho_{0k}+\epsilon^2\delta\rho_k(\epsilon x, \epsilon t),\\ \label{perturbation:v}
	v_k &=\epsilon^2 \delta v_k(\epsilon x,\epsilon t),
	\end{align}
	where $\epsilon$ is a small formal parameter. From here on, we limit ourselves to the case where $m$ is a scalar and common to all species. All of our analysis on the linear system (and the resultant perturbation scheme) extends to the case when each species has a corresponding distinct value for $m$, without any change in our conclusions. However, for the sake of simplicity of presentation we limit ourselves to a single common value for $m$. Moreover, we set that value to $m=1$ without loss of generality. 

	Substituting the form of the perturbation (\ref{perturbation:rho}-\ref{perturbation:v}) into the hydrodynamic equations and dropping terms of $\mathcal{O}(\epsilon^2)$ we obtain
	\begin{equation}
		\del_t\hspace{-2pt}
		\begin{pmatrix}
		\delta\rho
		\\
		\delta v
		\end{pmatrix}\hspace{-4pt}=-\del_x\hspace{1pt}
		\mathcal A
		\begin{pmatrix}
		\delta\rho\\
		\delta v
		\end{pmatrix},	
	\end{equation}
	where
	\begin{equation}
		 \mathcal A = \left( \begin{array}{cc} \mathbf{0}_{N\times N} & \rho \\ \alpha & \mathbf{0}_{N\times N} \end{array} \right), 
	\end{equation} with $\rho$  an $N\times N$ diagonal matrix with elements $\rho_{0k}>0$ and $\delta \rho\,,\delta v$ are the $N\times 1$ vectors for the perturbations in density and velocity. Real eigenvalues of the matrix $\mathcal{A}$ correspond to traveling wave solutions, either to right or left, depending on the sign of the eigenvalue. Complex eigenvalues, on the other hand, correspond to unstable exponentially growing modes. Evidently, for a stable background state we must choose the coupling matrix $\alpha$ such that all eigenvalues of $\mathcal A$ are real. 

	If all eigenvalues are indeed real, the solution consists of pulses propagating at speeds given by each of the eigenvalues. These speeds, referred to as sound speeds for the system, depend on the values of the coupling constants, background densities and mass of the condensate atoms. These may be readily measured in an experiment and compared to the theoretically predicted values. 
As we show in the subsequent section, for a specific model of the coupling matrix, the sound speeds are easily computed.

	\subsection{\label{sec:IIC}Spectral analysis of \texorpdfstring{$\mathcal A$}{TEXT}}
	As emphasized in the previous section, the eigenvalues of $\mathcal A$ play a crucial role in designing the multicomponent system. As we will see in the following section, both eigenvalues and eigenvectors play a key role in deriving equations governing the nonlinear dynamics of small amplitude pulses. The purpose of the present section is to obtain a full characterisation of the spectrum of $\mathcal A$. Indeed, we present arguments to compute eigenvalues, assure their reality, determine their multiplicity and calculate eigenvectors. As a consequence, we are able to determine the spectral decomposition of the $2N\times 2N$ matrix $\mathcal A$ as a function of the coupling coefficients $\alpha$. \vv{Our main assumption is that the coupling matrix is a real symmetric positive definite matrix. Under this assumption alone, Theorems \ref{thm:A:PosDef} and \ref{thm:A:vectors} characterize the spectrum (eigenvalues and eigenvectors) of $\mathcal A$ in terms of the spectrum of $\alpha$ and the background densities. It is this information that is needed to perform the reductive perturbation theory of the following section.}

	\vv{To obtain more information on the eigenvalues/eigenvectors of $\mathcal A$ (mulitiplicity of eigenvalues, the characteristic polynomial, assure reality and positivity of eigenvalues of $\alpha$, etc.) we make the following assumption for $\alpha$
	\begin{assumption}\label{model_assumption}
		\[\alpha_{ij} = \left\{ \begin{array}{ll} g_i, & i = j, \\ h, & i\neq j, \end{array} \right.\]
		where $g_i\,,h$ are all positive constants.
	\end{assumption}
	Theorems \ref{thm:alpha:PosDef}, \ref{thm:CharPoly}, \ref{thm:A:multiple} and \ref{thm:A:degen_vectors} invoke the above assumption. On the other hand, the arguments presented in the proof of these theorems can be adapted to alternate forms of the coupling matrix.  We present more detail than typical for the interested reader to adapt these arguments for alternate coupling matrices. }
 
	We begin with a sufficient condition to ensure real eigenvalues for $\mathcal A$. Note, that since $\mathcal A$ is not symmetric, we are not readily guaranteed real eigenvalues. 

	{\theorem{\label{thm:A:PosDef}If $\alpha$ is symmetric positive definite, then the eigenvalues of $\mathcal A$ are non-zero, real and come in pairs with opposite sign.}} 

	\begin{proof}
		Any eigenvalue $\lambda$ of $\mathcal A$ satisfies the characteristic polynomial $\mathrm{det}(\mathcal A - \lambda) = \mathrm{det}(\lambda^2 - \alpha \rho)$ where the equality follows from the block nature of $\mathcal A$. Hence, all $\lambda$ are real if and only if $\alpha\rho$ has positive eigenvalues. Since $\rho$ is a diagonal matrix with positive elements, we have $\alpha\rho = \rho^{-1/2}(\rho^{1/2}\alpha\rho^{1/2})\rho^{1/2}$. Hence $\alpha\rho$ is similar to $\rho^{1/2}\alpha\rho^{1/2}$ and therefore these matrices have the same eigenvalues. On the other hand, $\rho^{1/2}\alpha\rho^{1/2}$ is congruent to $\alpha$ and so these matrices have the same number of positive eigenvalues. Since $\alpha$ by assumption is positive definite, $\rho^{1/2}\alpha\rho^{1/2}$ has $n$ positive eigenvalues and consequently, $\alpha\rho$ has $N$ positive eigenvalues. Thus $\mathcal A$ has $N$ positive eigenvalues and $N$ negative eigenvalues.
	\end{proof}

	We now present a necessary condition for the coupling matrix to be positive definite. Thus if the condition is violated, $\alpha$ cannot be positive definite and we expect perturbations to the background to be unstable. To be specific, we have the following theorem.

	{\theorem{
	\vv{Let $\alpha$ satisfy Assumption \ref{model_assumption}.} Let the components of the system be ordered such that $g_i\leq g_{i+1}$, $i=1,2,\ldots$. Then $\alpha$ is a positive definite matrix implies $h<\sqrt{g_1g_2}$.  \label{thm:alpha:PosDef}
	}}
	
	\begin{proof} See Appendix \ref{p-thm:alpha:PosDef}. 
	\end{proof} \\

It is to be noted that the above is a necessary condition but not sufficient. A simple sufficient condition is that $h<g_1 $. To see this, let $x$ be any vector in $\mathbb{R}^N$. A straightforward computation shows $x^T\alpha x = \sum_i x_i^2 (g_i-h) + h \left(\sum_i x_i\right)^2$. If $h<g_1\Rightarrow h<\min g_i$, then $x^T\alpha x$ is automatically positive for any vector $x$ and so $\alpha$ is in fact positive definite. 

%

	{\cor{The theorem above assumed a particular ordering. However, the eigenvalues of the system do not depend on the ordering; different orderings being obtained as mere permutations of the same system of equations. Hence the conclusion of the theorem holds when $g_1,g_2$ are interpreted as the two smallest diagonal elements of $\alpha$.}}\\


	Assumption \ref{model_assumption} allows us to compute the characteristic polynomial of $\mathcal A$ in closed form. We first, however, introduce some notation. 
	
	{\definition{ 
	Let $\{\gamma_i\}_{i=1}^N$ be a list of real numbers and $C_{Nk}$ be the set of all possible ways to choose any $k$ of these $N$ real numbers. We denote by 
	$\Sym{\gamma_i}{k}$ the symmetric product 
	\begin{equation}
		 \Sym{\gamma_i}{k} = \sum_{\sigma\in C_{Nk}} \prod_{\gamma_j\in \sigma}\gamma_j
	\end{equation}
	}}

	The definition above gives symmetric products of $\gamma_i$. For example, $\Sym{\gamma_i}{n} = \prod_j \gamma_j $ and $\Sym{\gamma_i}{1} = \sum_{j} \gamma_j$. In the case of only three elements $\gamma_i,\ i=1,2,3$, then $\Sym{\gamma_i}{2} = \gamma_1\gamma_2 + \gamma_2\gamma_3 + \gamma_3\gamma_1$. 

	{\definition{
	We define 
	\begin{equation}
				\Sym{\gamma_i}{0} = 1, \quad \Sym{\gamma_i}{m} = 0,\ m>N
	\end{equation}
	}}

	Using the above notation,  we state the following theorem. 

	{\theorem{  
	The characteristic polynomial of $\mathcal A$ is \label{thm:CharPoly}
	\begin{align}\nonumber
	&\Sym{\frac{\rho_{0i} g_i - \lambda^2}{\rho_{0i} h}}{N} \\  \label{eqn:CharPoly}
	+& \sum_{k=2}^N (-1)^{k-1}(k-1) \Sym{\frac{\rho_{0i} g_i-\lambda^2}{\rho_{0i} h}}{N-k} =  0,
	\end{align} 
	\vv{when $\alpha$ satisfies Assumption \ref{model_assumption}.}
	}}

	The above theorem may be proved using the principle of induction; the base case $N=2$ is easily checked by hand. Though straightforward, the proof by induction is involved and employs some particular properties of the symmetric products we have defined. The details are presented in Appendix \ref{app:proof:charpoly}. 

	The characteristic polynomial of $\mathcal A$ is evidently a polynomial in $\lambda^2$ (as expected from Theorem \ref{thm:A:PosDef}). However, it is also a polynomial in $h$ as seen by multiplying the entire expression (\ref{eqn:CharPoly}) by $h^N$. Indeed, the characteristic polynomial is a polynomial in $\lambda^2$ with coefficients that are polynomials in $h$. Such an expression is called an algebraic curve. These objects are well studied amongst mathematicians, though we only require some very basic properties of such polynomials.

	Note for each $h$, there are $2N$ values of $\lambda$ that are roots of the characteristic polynomial (choosing $h$ suitably so that the values of $\lambda$ are all real). Since the eigenvalues of $\mathcal A$ represent the physical speeds of the small amplitude pulse-like initial perturbations to the base state, we are especially interested in knowing whether all sound speeds of the system are distinct. Distinct speeds correspond to pulses propagating such that eventually they do not interact. Consequently, determining whether the sound speeds are distinct is equivalent to determining whether the eigenvalues of $\mathcal A$ are distinct.

	An important property of algebraic curves is that either (a) for most values of $h$ (in particular, except for a finite number of isolated values of $h$), the roots $\lambda^2$ of (\ref{eqn:CharPoly}) are distinct, or (b) the number of distinct roots of (\ref{eqn:CharPoly}) is less than $2N$ for all values of $h$. 

	A polynomial under case (b) is said to be permanently degenerate. We will state necessary and sufficient conditions on the coupling matrix $\alpha$ such that case (b) holds. Moreover, we will also determine an analytic expression for the repeated eigenvalues. A multicomponent system prepared such that case (a) is true, namely, when (b) does not hold will almost surely have distinct sound speeds since to have repeated eigenvalues, very precise values of $h$ must be chosen. Any perturbation of these particular values for $h$ will immediately lead to distinct sound speeds. If the conditions for case (b) are not satisfied, then we will assume the eigenvalues are simple for a given $h$. In other words, the sound speeds are distinct. The following theorem states conditions for case (b).

	{\theorem{\vv{Suppose $\alpha$ satisfies Assumption \ref{model_assumption}.} The characteristic polynomial of $\mathcal A$ is permanently degenerate with a root of multiplicity $m-1$ if and only if $m$ pairs of $(\rho_{0i}g_i,\rho_{0i})$ are equal. Furthermore, if the common value of the $(\rho_{0i}g_i,g_i)$ pairs is denoted by $(\rho_0^* g^*,\rho_0^*)$, then $\pm\sqrt{\rho_0^*(g^*-h)}$ are the associated repeated eigenvalue of $\mathcal A$.\label{thm:A:multiple} }}\\

	\begin{proof}
		See Appendix \ref{app:proof:multiple}.
	\end{proof}
	{\cor{Some implications of the above theorem are,
	\begin{itemize}
	\item if two $(\rho_{0i}g_i,\rho_{0i})$ pairs are equal (to say $\rho_0^* g^*,\rho_0^*$), then $\pm\sqrt{\rho_0^* (g^* - h)}$ must be eigenvalues of $\mathcal A$.
    \item the above theorem is true for any repeated pair $(\rho_{0i}g_i,\rho_{0i})$ and thus holds for each repeated pair. 
    \item suppose another $m'$ pairs of $(\rho_{0i}g_i,\rho_{0i})$ were equal (and distinct from the first $m$ pairs). Then the characteristic polynomial takes the following form
    \begin{equation}
    	  \left(\frac{\rho^*_0 g^* -\lambda^2}{\rho^*_0 h} - 1\right)^{m-1}\left(\frac{\rho'_0 g' -\lambda^2}{\rho'_0 h} - 1\right)^{m'-1}\ \psi(\lambda^2)=0
    \end{equation}
    with
    \begin{align} 
    \psi(\lambda^2) &= \sum_{p=0}^{N-m-m'}\left[ \Sym{\frac{\rho_{0i} g_i -\lambda^2}{\rho_{0i} h}}{p} \ \right. \times \nonumber \\
    &\quad\quad\quad\: \left.  (-1)^{N-m-m'-p}f(\gamma,\gamma',p)\bigg.\right]
    \end{align}
    where 
    \begin{align} 
    f(\gamma,\gamma',p)) &= m(\gamma'-1) + m'(\gamma-1)\nonumber \\ &\quad  + (\gamma-1)(\gamma'-1)(1+N-m-m'-p)
    \end{align}
    and \begin{equation}
    	\gamma = \left(\frac{\rho^*_0 g^* - \lambda^2}{\rho^*_0 h}\right), \quad \gamma' = \left(\frac{\rho'_0 g' - \lambda^2}{\rho'_0 h}\right). 
    \end{equation}
	\end{itemize}}}

	Having fully characterized the eigenvalues of $\mathcal A$, we now proceed to investigate the eigenvectors. \vv{Our first result states that $\mathcal A$ is diagonalisable. Note this result \textit{does not} require Assumption \ref{model_assumption}.}

	{\theorem{  Assuming the matrix $\alpha$ is positive definite, the matrix $\mathcal A$ has $2N$ independent eigenvectors. In particular, the algebraic and geometric multiplicities of any eigenvalue of $\mathcal A$ are equal for all permissible values of $h$. In other words, the matrix $\mathcal A$ is diagonalisable.\label{thm:A:vectors}}}

	\begin{proof}
		The $N\times N$ matrix $\rho^{1/2}\alpha \rho^{1/2}$ is real, symmetric and positive definite whenever $\alpha$ is real, symmetric and positive definite. Hence there exist $N$ mutually orthogonal eigenvectors $u^i,\ i=1,2,\ldots N$ that span $\mathbb{R}^N$. Denote the eigenvalue associated with $u^i$ by $\lambda^2_i$ and let $q^i = \rho^{-1/2} u^i$. Then we have
    	\begin{equation}
    		 \alpha\rho\: q^i = \rho^{-1/2}\rho^{1/2}\alpha\rho^{1/2}\:u^i = \lambda^2_i \rho^{-1/2}u^i =  \lambda^2_i q^i, 
    	\end{equation} and hence $q^i$ is  an eigenvector of $\alpha\rho$ with eigenvalue $\lambda^2_i$. This is true for each $i$ and hence we have determined $N$ eigenvectors for $\alpha\rho$. Consequently, the matrix $\alpha\rho$ is diagonalisable: the algebraic and geometric multiplicities are equal. We claim each $q^i$ induces two $2N-$dimensional eigenvectors $v^i_{\pm}$ for the matrix $\mathcal A$ corresponding to eigenvalues $\pm\lambda_i$. Indeed, define 
    	\begin{equation}
    	v^i_{\pm} = \left[ \begin{array}{c} \dfrac{\pm1}{\lambda_i}\rho\:q^i  \vspace{5pt} \\ q^i \end{array} \right].
    	\end{equation}
    	It then follows that $\mathcal Av^i_{\pm} = \pm\lambda_i v^i_{\pm}$ where $\lambda_i$ is the positive root of $\lambda^2_i$, the associated eigenvalue for $\alpha\rho$. Let $Q$ represent the matrix with columns $q^i$ and $\Lambda$ be the diagonal matrix with diagonal elements $\lambda_i$. Then the matrix whose columns are eigenvectors of $\mathcal A$ is given by \begin{equation}
    	V = \left[\begin{array}{cc} \rho Q \Lambda^{-1} & - \rho Q \Lambda^{-1} \\ Q & Q \end{array} \right], 
    	\end{equation} with determinant 
    	\begin{align}
    		\det({V}) &= \det{\big(\rho Q\Lambda^{-1}Q - (-\rho Q\Lambda^{-1}) Q} \big),\\ 
    		&= \det{( 2 \rho Q\Lambda^{-1}Q)}\neq 0,
    	\end{align}
    	since $\rho,Q,\Lambda^{-1}$ are all invertible matrices. Thus the columns of $V$ are linearly independent and hence there are $2N$ independent eigenvectors for $\mathcal A$.
	\end{proof}
	{\cor{If $q^i$ is an eigenvector of $\alpha\rho$ with eigenvalue $\lambda^2_i$, then 
	\begin{equation}
		 v^i_{\pm} = \left[ \begin{array}{c} \pm \dfrac{1}{\lambda_i}\rho q^i  \vspace{5pt} \\ q^i \end{array} \right],
	\end{equation} is an eigenvector of $\mathcal A$ with eigenvalue $\pm\lambda_i$. Hence computing eigenvectors of $\mathcal A$ is equivalent to computing those of $\alpha \rho$.}}

	\vv{When $\alpha$ satisfies Assumption \ref{model_assumption} and the characteristic polynomial is permanently degenerate (has repeated eigenvalues for all values of $h$), Theorem \ref{thm:A:multiple} states the exact expression of these repeated eigenvalues and their multiplicity. The following theorem provides an exact form for the associated eigenvector in this case.}

	{\theorem{\label{thm:A:vec} \vv{Suppose $\alpha$ satisfies Assumption \ref{model_assumption} and} the matrix $\mathcal A$ \label{thm:A:degen_vectors} is permanently degenerate, \emph{i.e.} there is an eigenvalue of multiplicity $m>1$ for all suitable values of $h$. Let $(\rho^*_0g^*,\rho^*_0)$ represent the repeated pair resulting in the degenerate eigenvalue. A set of $m$ independent eigenvectors for these permanently degenerate eigenvalues $\lambda = \pm\sqrt{\rho^*_0(g^*_0-h)}$ of multiplicity $m$ are given by 
	\begin{equation}
		v^{(k)} = \left[ \begin{array}{c} \dfrac{\pm 1}{\sqrt{\rho^*_0(g^*_0-h)}}\rho q^{(k)}\vspace{5pt} \\ q^{(k)}\end{array}\right],
	\end{equation}
	where the $i-$th component of the $k-$th eigenvector is given by
	\begin{equation}
		(q^{(k)})^i =  \left\{ \begin{array}{ll} 1, & i=i_1,\\ -1, & i=i_{k+1},\\ 0, & \mbox{else,} \end{array}\right.
	\end{equation} for $k=1,2,\ldots m$. The indices $i_k,\: k=1,2,\ldots,m+1$ are such that the diagonal elements of $\alpha\rho$ at these locations $(\alpha\rho)_{i_ki_k}=\rho^*_0 g^*$.
	}}
	
	\begin{proof} See Appendix ~\ref{p-thm:A:vec}. 
	\end{proof}

	\subsubsection{Numerical method to compute eigenvalues and eigenvectors of \texorpdfstring{$\mathcal A$}{TEXT}}\label{sec:IIC:numerical}
	The previous theorems establish the reality of the eigenvalues of $\mathcal A$. Moreover, we have explicit formulae for the permanently repeated eigenvalues (if any) and their associated eigenvectors. It remains to investigate the eigenvalues and eigenvectors which are typically simple, \emph{i.e.} simple for most values of $h$. Although one could simply use a standard numerical solver to compute roots of the characteristic polynomial for various $h$, we present some further analytic results and a simple iterative procedure that determines both eigenvalues and eigenvectors. 

	From Theorem \ref{thm:A:multiple}, non-degenerate eigenvalues correspond to those $(\rho_{0i}g_i,\rho_{0i})$ pairs which do not repeat. Notice that when $h=0$, the characteristic polynomial (\ref{thm:CharPoly}) has roots $\lambda=\pm\sqrt{g_i\rho_{0i}}$. We conclude that at $h=0$, the non-repeating $\rho_{0i}g_i$ are simple eigenvalues. It is known that this behaviour, namely the simple nature of the eigenvalue, must persist at least for small $h$ \cite{kato2013perturbation}. The claim essentially follows from the implicit function theorem. We now present a scheme to compute both eigenvalues and eigenvectors as a function of $h$ that limit to a non-repeated $\rho_{0i},g_i$ when $h=0$.  This presentation employs the particular structure of the coupling matrix dictated by Assumption \ref{model_assumption}. 

	We work directly with the matrix $\alpha\rho$ instead of $\mathcal A$. Theorems \ref{thm:A:PosDef} and \ref{thm:A:vectors} readily allow us to translate spectra between the two matrices. An eigenvalue-eigenvector pair for $\alpha \rho$ satisfies the following equation \begin{equation}
		 \alpha\rho q = \lambda^2 q
	\end{equation} where $q$ is the eigenvector. When $h=0$, eigenvalues and eigenvectors are readily available: $\rho_{0i} g_i$ are eigenvalues with the canonical basis in $\mathbb{R}^N$ as the associated eigenvectors. We proceed to compute eigenvalues and eigenvectors for non-zero $h$ as follows.

	Let $q^{(k)} = e_k + h q_h^{(k)}$ be the eigenvector for $\alpha\rho$ associated with eigenvalue $\lambda^2 = \rho_{0k}g_k + h\mu_h^{(k)}$ for $h\neq 0$. Here we consider only those cases when $\rho_{0k},g_k$ is not repeated and $e_k$ represents the $k-$th canonical unit vector in $\mathbb{R}^N$. Under Assumption \ref{model_assumption} the matrix $\alpha\rho$ may be written as 
	\begin{equation}
		\alpha\rho = \alpha_0\rho + h \alpha_1\rho 
	\end{equation} where \begin{equation}
		 (\alpha_0\rho)_{ij} = \left\{ \begin{array}{ll} \rho_{0i}g_i, &i=j,\\ 0, & i\neq j,\end{array}\right.  ,\quad (\alpha_1)_{ij} =\left\{ \begin{array}{ll} 0, & i=j,\\ 1, & i\neq j. \end{array}\right. 
	\end{equation} Substituting for $\lambda^2$, $q$ and $\alpha\rho$, we have after some rearrangement
	\begin{equation}
		 (\alpha_0\rho-\rho_{0k}g_k) q_h^{(k)} = (\mu_h^{(k)}-\alpha_1\rho)(e_k + h q_h^{(k)})  
	\end{equation}

	Note that $\alpha_0\rho-\rho_{0k} g_k$ is diagonal and has a null-space: $e_k$. Thus we stipulate $q_h^{(k)}$ be orthogonal to $e_k$ and further require that the right-hand side of the above equation also be orthogonal to $e_k$. In more mathematical parlance, we are invoking the Fredholm alternative. The orthogonality condition for the right-hand side leads to 
	\begin{equation}
		\mu_h^{(k)} = \langle e_k,\alpha_1\rho e_k+h\alpha_1\rho q_h^{(k)}\rangle = h(\alpha_1\rho\: q_h^{(k)})_k.
	\end{equation}
	In other words, the correction to the eigenvalue is given by $h$ times the $k-$th component of $\alpha_1\rho \: q_h^{(k)}$. Substituting this expression into the equation for $q_h^{(k)}$ we obtain
	\begin{align}\label{eqn:iteration:eigenvector} (\alpha_0\rho-\rho_{0k}g_k) q_h^{(k)} = ( h(\alpha_1\rho\: q_h^{(k)})_k -\alpha_1\rho)(e_k + h q_h^{(k)}). \end{align}
	This equation is iteratively solved for $q_h^{(k)}$ since the left-hand side matrix is invertible (when $q_h^{(k)}$ is orthogonal to $e_k$). Notice that $\alpha_0\rho-\rho_{0k}g_k$ is diagonal and hence readily inverted. Once we converge to a $q_h^{(k)}$ that satisfies the above equation for some $h>0$, we evaluate the eigenvalue as
	\begin{align}
	    \lambda^2 = \rho_{0k}g_k + h^2 (\alpha_1\rho q_h^{(k)})_k.
	\end{align}
	We notice that the correction to the eigenvalue for $h>0$ is quadratic in $h$. If one expands $q_h^{(k)}$ in a power series of $h$, we find 
	\begin{equation}
		 \lambda^2 = \rho_{0k}g_k - h^2 \sideset{}{'}\sum_{j=1}^N \frac{\rho_{0j}\rho_{0k}}{\rho_{0j} g_j-\rho_{0k}g_k} + \ldots
	\end{equation}
	where the prime indicates the $k-$th term is skipped. 

	The iteration procedure described above may be justified by appealing to the implicit function theorem. Here we consider the correction to the eigenvector $q^{(k)}_h$ as a function of $h$. The requirements of the implicit function theorem hold at the point $h=0,q^{(k)}_h=0$, \emph{i.e.} the linearisation of expression \eqref{eqn:iteration:eigenvector} at $h=0,q^{(k)}_h=0$ leads to an invertible matrix. 

	At the outset, we do not know how large the radius of convergence (in $h$) of the resultant series is. However, given an eigenvalue-eigenvector pair for $\alpha\rho$ for $h\neq 0$, say $(e_k + hq^{(k)}_h,g_k\rho_k+h\mu_h^{(k)})$,  we may repeat the perturbation argument and restart the series around a non-zero value of $h$. Hence setting $q^{(k)}=e_k + h q^{(k)}_k + \delta hp^{(k)}_{\delta h}$ and $\lambda^2 = g_k\rho_{0k} + h\mu^{(k)}_h + \delta h\nu^{(k)}_{\delta h}$ we obtain the following equation for $p^{(k)}_{\delta h}$
	\begin{align}\nonumber
	\big(\: (\alpha_0+h\alpha_1)\rho-& g_k\rho_{0k}-h\mu_h^{(h)} \big) p_h^{(k)} = \\ \label{eqn:iteration:eigenvector:anyH}
	&( \nu_{\delta h}^{(k)} -\alpha_1\rho)(e_k + h q^{(k)}_h + \delta h q_{\delta h}^{(k)}),
	\end{align}
	where 
	\begin{align} \nu_{\delta h}^{(k)}=  \frac{\langle \rho q_h^{(k)},\alpha_1 \rho q_h^{(k)} \rangle + \delta h \langle \rho q_h^{(k)}, \alpha_1 \rho\: p_{\delta h}^{(k)} \rangle }{\langle \rho q_h^{(k)}, q_h^{(k)} \rangle + \delta h \langle \rho q_h^{(k)}, p_{\delta h}^{(k)}\rangle}. 
	\end{align}
	Once again by appealing to the implicit function theorem, one can establish that equation \eqref{eqn:iteration:eigenvector:anyH} can be solved for $p^{(k)}_{\delta h}$ for sufficiently small $\delta h$. By repeatedly using the above argument, we may obtain the eigenvector-eigenvalue for $\alpha \rho$ for all suitable $h$.

	\subsection{\label{sec:IID} Inhomogeneous linear dynamics}
	With an eye towards the calculations in the next section, we now discuss the solution procedure for inhomogeneous equations of the form
	\begin{align}\label{eqn:inhomo}
		\left(\partial_T + \mathcal A\partial_X\right) s = f,\quad \mathcal A = \left( \begin{array}{cc} 0 & \rho \\ \alpha & 0 \end{array}\right)
	\end{align}
	where we assume $f$ is a known $N\times 1$ vector valued function and we wish to determine the $N\times 1$ vector $s$. The main tool we employ is the Fredholm alternative.

	The Fredholm alternative is a statement on the solvability of linear equations. Consider a matrix equation $Lx = b$, where $L$ is a square matrix and $b$ is known. If $L$ is invertible, the solution is readily available: $x=L^{-1}b$. If however $L$ is not invertible, a necessary condition for a solution is that $b$ must be orthogonal to all $y$ such that $L^Ty=0$. This is the alternative. Note if $\langle y,b\rangle=0$ for all such $y$, we may have an infinite number of solutions. A unique solution can be obtained from an infinite possible set, if we also suppose $x$ is orthogonal to the null space of $L$.

	The above considerations for a matrix apply also to differential operators. Consider the equation $\partial_X\psi(X) = \zeta(X,T)$, where $\zeta$ is known. Clearly any constant is in the null space of the operator $\partial_X$. The adjoint of $\partial_X$ is $-\partial_X$ which also has constants as its null space. Hence we require the function $\zeta$ to be orthogonal to constants. To make these statements rigorous we need to state appropriate Hilbert spaces and inner products. We will avoid such technicalities presently.

	Coming back to equation (\ref{eqn:inhomo}), we recall that $\mathcal A$ has a spectral decomposition 
	\begin{align}
		\mathcal A &= V 
		\tilde\Lambda V^{-1},\quad V = \left( \begin{array}{cc} \rho Q\Lambda^{-1} & -\rho Q\Lambda^{-1} \\ Q & Q \end{array}\right), \nonumber \\
		\tilde \Lambda&=\left(\begin{array}{cc} \Lambda & 0 \\
		0 & -\Lambda \end{array} \right),
	\end{align} 
	where $\Lambda$ is a diagonal matrix with positive elements such that $\alpha\rho Q = Q\Lambda^2$. Substituting this into the inhomogeneous equation (\ref{eqn:inhomo}) we obtain
	\begin{align}
		\left( \partial_T + \tilde \Lambda\partial_X\right)V^{-1} s = V^{-1}f.
	\end{align}
	The null space of the differential operator on the left-hand side above consists of vectors of the form $\psi(x-\tilde\lambda_j t)e_j$, where $e_j$ is the $j-$th canonical unit vector in $\mathbb{R}^{2N}$, $\tilde\lambda_j$ is the $j-$th element along the diagonal of $\tilde\Lambda$ and $\psi$ is any function. The null space of the adjoint is the same. Thus the condition to solve the above equation for $V^{-1}s$ is that $\langle e_j, V^{-1}f \rangle$ should not be a function of $(X-\tilde \lambda_j T)$. In other words $\langle e_j, V^{-1}f \rangle$ should not be proportional to $\phi(X-\tilde\lambda_j T),$ for any function $\phi$. In the next section we will see how this analysis of the linear inhomogeneous equation serves us in deriving equations governing the slow evolution of perturbations to the multi-component system \eqref{mnls}.

	\section{\label{sec:III} Reductive Perturbation method for N-component NLS}
	In the previous section we analysed the linearised equations for the perturbations $\delta \rho_k,\delta v_k$ about the trivial state \eqref{eqn:linear:hydro}. The fully nonlinear equations for the perturbations (in hydrodynamic form) without any additional scaling are
	\begin{align}
\label{eq15}
	\left(\partial_t+\mathcal A\partial_x\right)\begin{pmatrix} \delta \rho \\ \delta v \end{pmatrix} &= -\partial_x \begin{pmatrix} \mathcal N_1(\delta \rho,\delta v) \\ \mathcal N_2(\delta\rho,\delta v)\end{pmatrix},
	\end{align}
	where \begin{align}
\label{eq16}
	(\mathcal N_1)_k &= \delta\rho_k \: \delta v_k,\\
\label{eq17}
	 (\mathcal N_2)_k &= \frac{\delta v_k^2}{2} -\left(\frac{\hbar^2}{2}\right) \frac{2(\rho_{0k}+\delta\rho_k)\delta\rho_k'' - \delta\rho_k'^2}{4(\rho_{0k}+\delta \rho_k)^2}.
	 \end{align}
	Let us rescale the variables so that $\del_t\to\epsilon\del_t$, $\del_x\to\epsilon\del_x$ and $\delta\rho,\delta v\to \epsilon^2\delta \rho,\epsilon^2\delta v$. This scaling is equivalent to assuming the following for the original physical variables
	\begin{align}
\label{rpert}
	\rho_k &= \rho_k^{(0)}	+ \epsilon^2 \delta \rho_k(\epsilon x,\epsilon t), \\
	v_k &= \epsilon^2 \delta v_k(\epsilon x,\epsilon t).
\label{vpert}
	\end{align}
	This leads to
	\begin{align}
	\left(\partial_T+\mathcal A\partial_X\right)\begin{pmatrix} \delta \rho \\ \delta v \end{pmatrix} &= -\epsilon^2 \partial_X \begin{pmatrix} \mathcal N_1(\delta\rho,\delta v) \\ \mathcal N_{2}(\delta\rho,\delta v,\epsilon)\end{pmatrix},
	\end{align}
	where $X=\epsilon x, T = \epsilon t$ and \begin{equation}
		\mathcal N_2(\delta \rho,\delta v,\epsilon^2) = \frac{\delta v_k^2}{2} -\left(\frac{\hbar^2}{2}\right) \frac{2(\rho_{0k}+\epsilon^2\delta\rho_k)\delta\rho_k'' - \epsilon^2\delta\rho_k'^2}{4(\rho_{0k}+\epsilon^2\delta \rho_k)^2},
	\end{equation}
	and $\delta\rho_k'' = \partial_X^2 \delta \rho_k$.

	We now solve the above equation perturbatively. Assuming an expansion in $\epsilon^2$ for the unknowns
	\begin{equation}
		\begin{pmatrix} \delta \rho \\ \delta v \end{pmatrix} = \begin{pmatrix}\delta\rho^{(0)} \\ \delta v^{(0)}\end{pmatrix}  + \epsilon^2 \begin{pmatrix}\delta\rho^{(1)} \\ \delta v^{(1)}\end{pmatrix},
	\end{equation}
	and substituting into the above equation we obtain to lowest order
	\begin{align}
\label{eq21}
		\left(\partial_T+\mathcal A\partial_X\right)\begin{pmatrix} \delta \rho^{(0)} \\ \delta v^{(0)} \end{pmatrix} &= \begin{pmatrix} 0 \\ 0 \end{pmatrix}.
	\end{align}
	Since $\mathcal A=V\tilde\Lambda V^{-1}$ is diagonalisable, this equation is equivalent to 
	\begin{align}
		\left(\partial_T+\tilde\Lambda \partial_X\right)V^{-1}\begin{pmatrix} \delta \rho^{(0)} \\ \delta v^{(0)} \end{pmatrix} &= \begin{pmatrix} 0 \\ 0\end{pmatrix},
	\end{align}
	which has a solution 
	\begin{equation}
		V^{-1}\begin{pmatrix} \delta \rho^{(0)} \\ \delta v^{(0)} \end{pmatrix} = f^{(0)}_j(X-\tilde\lambda_jT)e_j,
	\end{equation}
	or 
	\begin{equation}
		\begin{pmatrix} \delta \rho^{(0)} \\ \delta v^{(0)} \end{pmatrix} = f^{(0)}_j(X-\tilde\lambda_jT)Ve_j,
	\end{equation}
	where $f^{(0)}_j(\xi)$ is any function, $e_j$ is the unit vector in $\mathbb{R}^{2N}$ and $\tilde \lambda_j$ is any of the eigenvalues of $\mathcal A$. As common in the method of multiple scales, we will assume $f^{(0)}_j$ depends on $X-\tilde \lambda T$ as well as a new slow time scale $\tau=\epsilon^2 T=\epsilon^3 t$. Hence 
	\begin{equation}
		 \begin{pmatrix} \delta \rho^{(0)} \\ \delta v^{(0)} \end{pmatrix} = f^{(0)}_j(X-\tilde\lambda_jT,\tau)Ve_j.
	\end{equation}

	The equations at order $\epsilon^2$ are then given by
	\begin{align}\nonumber
		\left(\partial_T+\mathcal A\partial_X\right)\begin{pmatrix} \delta \rho^{(1)} \\ \delta v^{(1)} \end{pmatrix} &= -\partial_\tau \begin{pmatrix} \delta \rho^{(0)} \\ \delta v^{(0)} \end{pmatrix}\\
		& - \partial_X \begin{pmatrix} \mathcal N_1(\delta \rho^{(0)},\delta v^{(0)}) \\ \mathcal N_2(\delta \rho^{(0)},\delta v^{(0)},0) \end{pmatrix},
	\end{align}
	which is equivalent to 
	\begin{align}\nonumber
		\left(\partial_T+\tilde\Lambda \partial_X\right)V^{-1}&\begin{pmatrix} \delta \rho^{(1)} \\ \delta v^{(1)} \end{pmatrix} = -\partial_\tau V^{-1}\begin{pmatrix} \delta \rho^{(0)} \\ \delta v^{(0)} \end{pmatrix}\\
		& - \partial_X V^{-1} \begin{pmatrix} \mathcal N_1(\delta \rho^{(0)},\delta v^{(0)}) \\ \mathcal N_2(\delta \rho^{(0)},\delta v^{(0)},0) \end{pmatrix}.
	\end{align}
	Notice this is a linear inhomogeneous equation for the order $\epsilon^2$ correction to $\delta\rho,\delta v$. The right-hand side is essentially a known function since every term on the right hand side can be written in terms of $f^{(0)}_j(X-\tilde \lambda_jT,\tau)$. Moreover, the adjoint of the linear operator on the left-hand side has a null space: precisely those functions of the form $\psi(X-\tilde\lambda_jT)e_j$. From the Fredholm alternative, the right-hand side should be orthogonal to this null space. Notice all terms on the right are of the form $\psi(X-\tilde\lambda_j T)$. Hence we have the solvability condition
	\begin{align*} 
	\left\langle e_j,  -\partial_\tau V^{-1}\begin{pmatrix} \delta \rho^{(0)} \\ \delta v^{(0)} \end{pmatrix} - \partial_X V^{-1} \begin{pmatrix} \mathcal N_1(\delta \rho^{(0)},\delta v^{(0)}) \\ \mathcal N_2(\delta \rho^{(0)},\delta v^{(0)},0) \end{pmatrix}\right\rangle \\ \quad = 0,
	\end{align*}
	where $e_j$ being the $j-$th unit vector of the $2N\times 2N$ identity matrix. The above equation can be simplified using the expression for the zeroth order solution to
	\begin{equation}
		 \left\langle e_j,  -\partial_\tau f^{(0)}_je_j - \partial_X V^{-1} \begin{pmatrix} \mathcal N_1(\delta \rho^{(0)},\delta v^{(0)}) \\ \mathcal N_2(\delta \rho^{(0)},\delta v^{(0)},0) \end{pmatrix}\right\rangle = 0,
	\end{equation}
	or in other words 
	\begin{equation}\label{eqn:KdV:2N}
		\partial_\tau f^{(0)}_j +  \left\langle e_j, \partial_X V^{-1} \begin{pmatrix} \mathcal N_1(\delta \rho^{(0)},\delta v^{(0)}) \\ \mathcal N_2(\delta \rho^{(0)},\delta v^{(0)},0) \end{pmatrix}\right\rangle=0.
	\end{equation}
	Rewriting $\mathcal N_1,\mathcal N_2$ entirely in terms of $f^{(0)}_j$ we have the required KdV equation. \vv{This is true for each $j$ and hence we have $2N$ KdV equations, $N$ of which correspond to perturbations traveling to the right and $N$ of which correspond to perturbations traveling to the left. If all sound speeds (i.e. eigenvalues $\lambda_j$ of $\mathcal A$) are distinct, these equations are uncoupled since the  terms $\mathcal N_1,\mathcal N_2$ are given entirely in terms of the profile $f^{(0)}_j$.} Physically this corresponds to moving into different traveling frames centered around each pulse. 

	To summarize, if $V$ is the eigenvector matrix of $\mathcal A$, namely $\mathcal A = V\tilde \Lambda V^{-1}$ then the density and velocity vectors are given by 
	\begin{align}\label{relation2}
		\begin{pmatrix}
		\vec\rho\\
		\vec v
		\end{pmatrix} 
		= 
		\begin{pmatrix}
		\vec\rho_0 \\
		0
		\end{pmatrix}
		+ \epsilon^2 f^{(0)}_j(\epsilon x-\lambda_j \epsilon t,\epsilon^3 t) V e_j + \mathcal{O}(\epsilon^4),
	\end{align}
	where $\lambda_j$ is eigenvalue associated with the eigenvector $Ve_j$ ($e_j$ being the $j-$th unit vector of the $2N\times 2N$ identity matrix).

	\vv{The upshot of the above analysis is a reduction of the dynamics of the coupled GPE (\ref{mnls}) in terms of $2N$ KdV equations. Physically speaking, a generic perturbation to the background densities, on the shortest timescale evolves according to (\ref{mnls}) in such a way so as to give rise to $2N$ small-amplitude waves traveling at the sound speeds (given by the eigenvalues of $\mathcal A$). On a longer timescale, the waves evolve according to the corresponding KdV equation given in (\ref{eqn:KdV:2N}). Hence a generic perturbation to the background state resolves into $2N$ waves evolving according to KdV ($N$ going to the right; $N$ going to the left). }

	\subsection{Coupled KdV equations (non-distinct speeds)}
	In the derivation presented in the previous section we assumed a solution to the homogeneous problem that depended only on one profile $f^{(0)}_j$. This is however not the most general solution. Indeed one may have well assumed \begin{equation}
		V^{-1}\begin{pmatrix}\delta\rho^{(0)}\\ \delta v^{(0)} \end{pmatrix} = \sum_j f^{(0)}_j(X-\lambda_j T,\epsilon^2 T)e_j.
	\end{equation} 	Evidently now \begin{equation}
		\partial_X V^{-1} \begin{pmatrix} \mathcal N_1(\delta \rho^{(0)},\delta v^{(0)}) \\ \mathcal N_2(\delta \rho^{(0)},\delta v^{(0)},0) \end{pmatrix},
	\end{equation} contains functions of all $X-\lambda_j T$ leading to what one may consider to be a coupled system of KdV. However, it must be noted, that when we project onto $e_j$ to obtain the equation of evolution for $f^{(0)}_j(X-\lambda_j T)$ we only retain those terms for the right-hand side which are functions of $X-\lambda_jT$ alone (and not products of functions of multiple $X-\lambda_j T)$. Hence once again we end up with uncoupled equations.

	Note the above argument fails when $\mathcal A$ has repeated eigenvalues. This is precisely why we determined necessary and sufficient conditions for simple non-repeating eigenvalues. In the case of repeated eigenvalues however, the zeroth order solution is given as 
	\begin{equation}
		V^{-1}\begin{pmatrix}\delta\rho^{(0)}\\ \delta v^{(0)} \end{pmatrix} = \sum_k f^{(0)}_k(X-\lambda T,\epsilon^2 T)e_k,
	\end{equation}where now the sum only extends over those vectors $Ve_k$ which correspond to the same eigenvalue $\lambda$. All the functions $f_j^{(0)}$ depend on the same spatial variable $\xi = X-\lambda T$ and slow time scale $\tau =\epsilon^2 T$. In the case of repeated eigenvalues, we necessarily obtain a coupled system of KdV equations; the number of equations is equal to the multiplicity of the eigenvalue. The dynamics of the coupled KdV equations arising out of repeated eigenvalues for $\mathcal{A}$ will be discussed in a future paper. 

	A final scenario that may also lead to coupled equations is when the eigenvalues are close together, indeed when $|\lambda_{j+1}-\lambda_j|<\epsilon^2$. In such a case, although the asymptotic behaviour of such a system is described by two uncoupled KdV equations (since they correspond to two different traveling frames of reference), due to the small difference in sound speeds, the dynamics may appear to be coupled even on the longer time scale for KdV-type equations. Note however, the resultant coupled system will typically have different coefficients than the one corresponding to repeated eigenvalues (when the sound speeds are exactly the same) since the associated eigenvectors are different in either case.

	\subsection{Some useful relations}\label{sec:useful_relations}
	Suppose $Q$ is the matrix of eigenvectors of $\alpha\rho$ with eigenvalues given by the diagonal matrix $\Lambda^2$. In other words $\alpha\rho Q = Q \Lambda^2$. All diagonal entries of $\Lambda^2$ are positive (see proof of Theorem \ref{thm:A:vectors}). Then the matrix of eigenvectors for $\mathcal A$ is given by \begin{align}\label{eq:vmat}
 V = \left(\begin{array}{cc} \rho Q \Lambda^{-1} & - \rho Q \Lambda^{-1} \\ Q & Q \end{array}\right),\end{align} with inverse  
	\begin{align} V^{-1} = \frac 1 2 \left(\begin{array}{cc} \Lambda Q^{-1}\rho^{-1} & Q^{-1} \\ -\Lambda Q^{-1}\rho^{-1} & Q^{-1} \end{array}\right). \end{align}
	For the purposes of deriving KdV, the relevant matrix is $V$ and $(V^{-1})^T$. It turns out, one may express $(V^{-1})^T$ explicitly in terms of $\rho,Q,\Lambda$. Indeed one has
	\begin{align} \label{eq:vinvtr}
	(V^{-1})^T = \frac 1 2 \left(\begin{array}{cc} Q L^{-1} \Lambda & - QL^{-1}\Lambda \\ \rho QL^{-1} & \rho QL^{-1} \end{array}\right), \end{align}
	where $L=Q^T\rho Q$. Moreover $L$ is a diagonal matrix with positive elements and hence $L^{-1}$ is readily computed. Note that $L=Q^T\rho Q$ implies $I=L^{-1}Q^T\rho Q$ and hence $Q^{-1}=L^{-1}Q^T\rho$.

	The above statements expressing $Q^{-1}$ in terms of $Q$ and $\rho$ are explained as follows. The matrix $\rho^{1/2}\alpha\rho^{1/2}$ is real symmetric and positive definite (assuming $\alpha$ is symmetric positive definite). As a result there exists a orthonormal matrix $U$ which is the eigenvector matrix of $\rho^{1/2}\alpha\rho^{1/2}$. Then 
	\begin{equation}
		\rho^{1/2}\alpha\rho^{1/2}U = U\mu^2 \Rightarrow \alpha\rho \rho^{-1/2}U = \rho^{-1/2}U\mu^2.
	\end{equation}
	But then we have $\rho^{-1/2}U$ is also an eigenvector matrix of $\alpha \rho$ and $\Lambda^2 = \mu^2$. This means $\rho^{-1/2}UM = Q$ where $M$ is a real diagonal matrix. In other words, the columns of $Q$ are parallel to columns of $\rho^{-1/2}U$. Expressing $U$ in terms of $Q,M,\rho^{1/2}$ and substituting in $UU^T=I$, $I$ being the $N\times N$ identity matrix, leads to $M^2=Q^T\rho Q$. We then define $L=M^2$, which is the matrix that appears in \eqref{eq:vinvtr}.

	We also note that if $D_1,D_2$ are diagonal matrices, then elements of $D_1 \alpha D_2$ are given by $(D_1\alpha D_2)_{ij} = (D_1)_i \alpha_{ij} (D_2)_j$ where $(D_1)_i,(D_2)_j$ denote the $i-$th and $j-$th diagonal entry of $D_1,D_2$ respectively. Hence any column of $V$ and $(V^{-1})^T$ is readily computed once the relevant column of $Q$ is determined. We recall Theorem \ref{thm:A:degen_vectors} and the procedure detailed in subsection \autoref{sec:IIC:numerical} allow us to compute a column of $Q$ independently of other columns. Of course, standard libraries provide all eigenvalues and eigenvectors simultaneously.

	With these definitions the coefficients of the respective KdV equations are obtained in a straightforward manner by considering the relevant column of $(V^{-1})^T$ for 
 	\begin{equation}
 		\partial_\tau f^{(0)}_j +  \left\langle e_j, \partial_X V^{-1} \begin{pmatrix} \mathcal N_1(\delta \rho^{(0)},\delta v^{(0)}) \\ \mathcal N_2(\delta \rho^{(0)},\delta v^{(0)},0) \end{pmatrix}\right\rangle=0,
 	\end{equation}
 	is equivalent to
 	\begin{align} \label{eq:kdv}
\partial_\tau f^{(0)}_j +  \left\langle (V^{-1})^Te_j, \partial_X \begin{pmatrix} \mathcal N_1(\delta \rho^{(0)},\delta v^{(0)}) \\ \mathcal N_2(\delta \rho^{(0)},\delta v^{(0)},0) \end{pmatrix}\right\rangle=0,\end{align}
 	where 
 	\begin{align} \label{relation1}
 	\begin{pmatrix}\delta\rho^{(0)}\\ \delta v^{(0)} \end{pmatrix} =  f^{(0)}_j(X - \lambda_j T,\tau)V e_j,\quad j=1,2,\ldots 2N.
 	\end{align}

 	\vv{We re-emphasize there are in total $2N$ KdV equations in (\ref{eq:kdv}).}

 	\subsection{A special case: \vv{KdV with zero nonlinearity}}

 	The matrix $\alpha$ represents the coupling between the different species and the matrix $\rho$ represents the trivial background states for the different species. Let us consider a case when two self-couplings (diagonal elements of $\alpha$) and their corresponding background states (the respective diagonal elements of $\rho$) are equal. In other words we assume $g_{i_1}=g_{i_2}=g^*$ and $\rho_{0i_1}=\rho_{0i_2}=\rho^*_0$ for some indices $i_1,i_2$. From Theorem \ref{thm:A:multiple} we are guaranteed that $\pm\sqrt{\rho^*_0(g^*-h)}$ are eigenvalues of $\mathcal{A}$. If more than two self-coupling--density pairs are equal, then the eigenvalues will have a multiplicity greater than one. Higher order multiplicities will be the focus of a future work and here we consider only the case of simple eigenvalues $\pm\sqrt{\rho^*_0(g^*-h)}$. We also limit the present discussion to the eigenvalue corresponding to waves traveling to the right. The analysis in this section extends similarly to the one traveling to the left.

 	Using Theorem \ref{thm:A:degen_vectors} we also know the exact form of the eigenvector associated with the eigenvalue $\sqrt{\rho^*_0(g^*-h)}$. Indeed it is  
 	\begin{equation}
 		v = \left[ \begin{array}{c} \dfrac{\rho}{\sqrt{\rho^*_0(g^*-h)}} q\vspace{5pt} \\ q\end{array}\right],
 	\end{equation}
	where $\rho$ is the matrix with diagonal elements $\rho_{0i}$ and the $i-$th component of $q$ is given by
	\begin{equation}
		q_i =  \left\{ \begin{array}{ll} 1, & i=i_1,\\ -1, & i=i_{2},\\ 0, & \mbox{else,} \end{array}\right.
	\end{equation}
	To obtain the relevant coefficients of KdV for this case, we need the relevant column of $(V^{-1})^T$ which is
	\begin{equation}
		\tilde v = \frac{1}{2} \begin{pmatrix}  \frac{\sqrt{\rho^*_0(g^*-h)}}{l}q \vspace{5pt} \\ \frac{\rho^*_0}{l} q  \end{pmatrix},
	\end{equation}
	where $l$ is the element of  $L=Q^T\rho Q$ corresponding to the vector $q$. Since $L$ is diagonal and $L=(\rho^{1/2}Q)^T\rho^{1/2}Q$ then $l = \sum_i \rho_{0i}q^2_i = 2\rho^*_0$. Thus 
	\begin{equation}
		\tilde v = \begin{pmatrix}  \frac{\sqrt{\rho^*_0(g^*-h)}}{4\rho^*_0} q \vspace{5pt} \\ \frac 1 4 q\end{pmatrix} 
	\end{equation}
	Then the KdV equation is given by 
	\begin{align}
		f_\tau + \left\langle \tilde v , \partial_X \begin{pmatrix} \mathcal{N}_1\left(\dfrac{\rho}{\sqrt{\rho^*_0(g^*-h)}} q,q\right) \vspace{2pt} \\ \mathcal{N}_2\left(\dfrac{\rho}{\sqrt{\rho^*_0(g^*-h)}} q,q,0\right) \end{pmatrix} \right\rangle = 0,
	\end{align}
	which upon simplifying is
	\begin{align}
		f_\tau - \frac{\hbar^2}{8\sqrt{\rho^*_0(g^*-h)}}f''' = 0,
	\end{align}
	where $f'''$ denotes the third-derivative with respect to the profile variable $X-\sqrt{(\rho^*_0(g^*-h)}\,T$. Remarkably this equation is linear and thus readily solved using Fourier transform techniques.

	\section{\label{sec:IV} Few component GPE/NLS and associated KdV and other findings}
	In this section we present explicit results for two specific cases, that of two and three component coupled systems. The reality of the eigenvalues (positivity of elements of $\Lambda^2$) ensures that all coefficients in the final KdV equation are well-defined real numbers.
	\subsection{\texorpdfstring{$N=2$}{N=2} case}
The coupled NLS are as follows
	\begin{subequations}\label{nls}
		\begin{gather}
			i\hbar\frac{\del \psi_1}{\del t}=-\frac{\hbar^2}{2m}\frac{\del^2\psi_1}{\del x^2}+g_1|\psi_1|^2 \psi_1+h|\psi_2|^2 \psi_1\\
			i\hbar\frac{\del \psi_2}{\del t}=-\frac{\hbar^2}{2m}\frac{\del^2\psi_2}{\del x^2}+h|\psi_1|^2 \psi_2+g_2|\psi_2|^2 \psi_2
		\end{gather}
	\end{subequations} 

In other words, the $\alpha$ matrix (under the assumption $m=1$; see text below Assumption \ref{model_assumption}) is given by 

\begin{equation}
\alpha = 
\begin{pmatrix}
	g_1  & h \\
	h & g_2
	\end{pmatrix}
\end{equation}

The above equations, \eqref{nls} can be written in a hydrodynamic form for a perturbation of a trivial state as given in Eqs.~\eqref{eq15} to \eqref{eq17} where $k = {1,2}$. We perform the perturbation series (\cref{rpert,vpert}) to arrive at Eq.~\eqref{eq21}
 which is the equation for the lowest order where the $\mathcal{A}$ is given by, 
\begin{equation}
\mathcal{A}_2=
\begin{pmatrix}
	0 & 0& \rho_{01} & 0\\
	0 & 0 & 0 &  \rho_{02}\\
	g_1&h&0&0\\
	h&g_2&0&0
	\end{pmatrix}
\end{equation}
where the subscript in $\mathcal{A}_2$ means that we are dealing with the two component case. 
The eigenvalues of the matrix $\alpha \rho$ are the diagonal elements of 
\begin{equation}
\Lambda ^2= 
\begin{pmatrix}
	\dfrac{A+B}{2} & 0\\
	0& \dfrac{A-B}{2}     
	\end{pmatrix}
\end{equation}
and eigenvectors are respectively, the columns of, 

\begin{equation}
Q= 
\begin{pmatrix}
	\dfrac{C+B}{2 h\rho_{01}} &  \dfrac{C-B}{2 h\rho_{01}} \vspace{3pt} \\ 
	1 &   1
	\end{pmatrix}
\end{equation}
where, $A,B,C$ are given by, 

\begin{subequations}
	\begin{gather}
	A = g_1\rho_{01}+g_2\rho_{02}\\
	C = g_1\rho_{01}-g_2\rho_{02}\\
	B = \sqrt{g_1^2 \rho_{01}^2 - 2 g_1 g_2 \rho_{01} \rho_{02} + 
		4 h^2 \rho_{01} \rho_{02} + g_2^2 \rho_{02}^2}
	\end{gather}
	\end{subequations}

Note that the eigenvalues (diagonal  elements of $\Lambda^2$) are both positive under the assumption $h<\sqrt{g_1g_2}$. This condition on $h$ is indeed not only necessary (Theorem \ref{thm:alpha:PosDef}) but also sufficient for the two component case. Next we employ the relations of Section \ref{sec:useful_relations}, using the definitions of $Q$ and $\Lambda$ given above, to  determine the coefficients of the respective KdV equations. Specifically, we determine the eigenvector matrix of $\mathcal{A}_2$ using \eqref{eq:vmat}. Similarly, we may compute $(V^{-1})^T$ using \eqref{eq:vinvtr}. In the following, we only present the results for the right chiral sector \emph{i.e.} two positive eigenvalues (sound speeds). 
Eq.~\ref{eq:kdv} for the two component species case (i.e., $N=2$) explicitly reads as follows for the two eigenvalues $\lambda = \sqrt{\dfrac{A\pm B}{2}}$ respectively,

\begin{eqnarray}
\partial_\tau f_j  + B_j f_j f_j^{\prime} +A_j f_j^{\prime\prime\prime}=  0, \quad j=1,2.\label{eq:kdvN2}
\end{eqnarray}
where,
\begin{eqnarray}
A_1 &=& -\frac{\hbar^2}{4\sqrt{2}} \frac{1}{\sqrt{A+B}} \\
B_1 &=&   \frac{3}{8hB\rho_{01}} \big[  (C+B)^2  +2 h (B-C) \rho_{01} \big] \\
A_2 &=&  -\frac{\hbar^2}{4\sqrt{2}} \frac{1}{\sqrt{A-B}}   \\
B_2 &=&  - \frac{3}{8hB\rho_{01}} \big[  (C-B)^2  -2 h (C+B) \rho_{01} \big] 
\end{eqnarray}

Here, $A_1, B_1$ are the KdV coefficients for the largest positive eigenvalue, $\lambda_1 = \sqrt{\frac{A+ B}{2}}$ and  $A_2, B_2$  are the  KdV coefficients for the second largest positive eigenvalue, $\lambda_2 = \sqrt{\frac{A- B}{2}}$. Here $f'''_j$ denotes the third-derivative with respect to the profile variable $X-\lambda_jT$. 

\vv{Equations (\ref{eq:kdvN2}) give the dynamics of perturbations to the background state in reference frames moving to the right with with speeds $\sqrt{(A\pm B)/2}$. Needless to say, there are two KdV equations for the other chiral sector namely for perturbations moving to the left with speeds $-\sqrt{(A\pm B)/2}$. The KdV equations for perturbations traveling to the left are obtained by setting $A_j\to -A_j$ in \eqref{eq:kdvN2}. Hence there are in total four KdV equations. \autoref{sec:V} contains the brute force numerical comparison between above KdV equation and the $N=2$ NLS case. }

\subsection{ \texorpdfstring{$N=3$}{N=3} case}
The $N=3$ case poses an interesting scenario. In general, for arbitrary $g_1, g_2, g_3$ and $\rho_{01}, \rho_{02},\rho_{03}$ (all different) the eigenvalues are very cumbersome but with our prescription outlined in the previous sections one can explicitly write it down. Below, we describe the situation when $g_1 = g_2$ and $\rho_{01} = \rho_{02}$. Here the eigenvalues and eigenvectors are still different. However, one of the eigenvalues and its corresponding eigenvector takes a particularly simple form. 

The three-coupled NLS are as follows
	\begin{subequations}\label{3nls}
		\begin{gather}
			i\hbar\frac{\del \psi_1}{\del t}=-\frac{\hbar^2}{2m}\frac{\del^2\psi_1}{\del x^2}+g_1|\psi_1|^2 \psi_1+h|\psi_2|^2 \psi_1 +h|\psi_3|^2 \psi_1 \\
			i\hbar\frac{\del \psi_2}{\del t}=-\frac{\hbar^2}{2m}\frac{\del^2\psi_2}{\del x^2}+h|\psi_1|^2 \psi_2+g_1|\psi_2|^2 \psi_2 + h|\psi_3|^2 \psi_2 \\
i\hbar\frac{\del \psi_3}{\del t}=-\frac{\hbar^2}{2m}\frac{\del^2\psi_2}{\del x^2}+h|\psi_1|^2 \psi_3+h|\psi_2|^2 \psi_3 + g_3|\psi_3|^2 \psi_3
		\end{gather}
	\end{subequations} 
In other words, the $\alpha$ matrix (with $m=1$) is given by 
\begin{equation}
\alpha = 
\begin{pmatrix}
	g_1  & h & h \\
	h & g_1 & h \\
         h & h & g_3
	\end{pmatrix}
\end{equation}
Correspondingly, we get 
\begin{equation}
\mathcal{A}_3=
\begin{pmatrix}
	0 & 0& 0& \rho_{01} & 0 & 0  \\
	0 & 0 & 0 & 0&  \rho_{01} & 0 \\
	0&0&0&0 & 0 & \rho_{03}\\
	g_1&h&h&0 & 0 & 0\\
	h&g_1&h&0&0&0 \\
         h&h&g_3&0&0&0
	\end{pmatrix}
\end{equation}
The eigenvalues of the matrix $\alpha \rho$ are the diagonal elements of 
\begin{equation}\label{eqn:lambda2:N3}
\Lambda ^2= 
\begin{pmatrix}
(g_1-h) \rho_{01} &0&0\\
	0&\frac{X+Y}{2} & 0\\
	0& 0& \frac{X-Y}{2}     
	\end{pmatrix}
\end{equation}
where we define,
\begin{eqnarray}
X&=& g_1 \rho_{01} + h \rho_{01} +g_3 \rho_{03}  \\
Y&=&  \Big[ (g_1+h)^2 \rho_{01}^2 -2 \big[ g_1 g_3+ h (g_3-4h) \big] \rho_{01}\rho_{03}
\nonumber \\ &+& g_3^2 \rho_{03}^2  \Big]^{\frac{1}{2}}
\end{eqnarray}
\vv{Corresponding to each of the three positive eigenvalues $\lambda_j$ of $\mathcal A$ (diagonal elements of $\Lambda$ in (\ref{eqn:lambda2:N3})), we obtain a KdV equation of the form}
\begin{eqnarray}
\label{lkdv}
\partial_\tau f_j  + B_j f_j f_j^{\prime} +A_j f_j^{\prime\prime\prime}=  0, \quad j=1,2,3.
\end{eqnarray}
Remarkably the corresponding KdV equation for the first eigenvalue $\lambda_1 = \sqrt{(g_1-h) \rho_{01}}$ is linear. Indeed
\begin{eqnarray}
A_1 &=& -\frac{\hbar^2}{8 \sqrt{(g_1-h)\rho_{01}}} \\
B_1 &=&   0 
\end{eqnarray}
For the other two eigenvalues, i.e., $\lambda_2= \sqrt{\frac{X-Y}{2} }$ and $\lambda_3= \sqrt{\frac{X+Y}{2} }$, we get $A_2, B_2,A_3,B_3$ as follows, 
\begin{eqnarray}
A_2 &=&  -\frac{\hbar^2}{4\sqrt{2}} \frac{1}{\sqrt{X-Y}} \\
B_2 &=& \frac{3\big[  2(Y-Z)^3 \rho_{01} + (W+Y)^3 \rho_{03} \big] }{4(W+Y)(Y-Z)^2 \rho_{01}+2 (W+Y)^3 \rho_{03}} \\
A_3 &=&  -\frac{\hbar^2}{4\sqrt{2}} \frac{1}{\sqrt{X+Y}} \\
B_3 &=& \frac{3\big[  -2(Y+Z)^3 \rho_{01} + (W-Y)^3 \rho_{03} \big] }{4(W-Y)(Y+Z)^2 \rho_{01}+2 (W-Y)^3 \rho_{03}} 
\end{eqnarray}
where $Z\,,W$ are given by, 
\begin{eqnarray}
Z &=& (g_1+h) \rho_{01} - g_3 \rho_{03} +2 h \rho_{03} \\ 
W &=& g_1 \rho_{01} -3 h \rho_{01} -g_3 \rho_{03}
\end{eqnarray}
In other words, the two eigenvalues other than $\lambda_1 = \sqrt{(g_1-h) \rho_{01}}$ have corresponding KdV equations for the nonlinear problem. 

\vv{We emphasize once more that equations (\ref{lkdv}) give the dynamics of perturbations that travel in the positive direction with speeds $\lambda_j$. There are three KdV equations for the other chiral sector namely, perturbations traveling to the left with speeds $-\lambda_j$. These KdV equations are obtained by setting $A_j\to -A_j$ in \eqref{lkdv}. Hence there are in total six KdV equations.}

\section{\label{sec:V} Numerical Results for two component case}
Following the results obtained in Sec. \ref{sec:IV} for $N=2$, in this section, we explore through numerical simulations the comparison between the obtained KdV and the coupled NLS. \vv{From our asymptotic analysis in that section, recall that a generic perturbation to the constant background density resolves into $2N$ weakly-nonlinear waves that each evolve according to KdV dynamics. To facilitate the comparison between NLS and KdV dynamics we choose specific initial conditions that generate perturbations which evolve according to only one of the KdV equations.}

Because of the integrability of the KdV \eqref{eq:kdvN2}, a soliton (or solitary wave profile) solution  is chosen as  a platform for comparison. Due to the special nature of these profiles, i.e. their ability to propagate and retain their structure without breaking by carefully balancing the effects of nonlinearity and dispersion, they also serve as a check on the numerics. 

Additionally, the validity of the simulations is also ensured by checking the conservation of wavefunction density and the Hamiltonian. The solitary wave solution is obtained for the KdV \eqref{eq:kdvN2} which is moving to the right with a velocity of $\mathcal{V}$ 
\begin{equation}\label{eqn:initkdv}
		f_j(\xi,\tau)=\frac{3\mathcal{V}A_j}{B_j}\text{sech}^{2}\bigg[\frac{\sqrt{\mathcal{V}}}{2}(\xi-A_j\mathcal{V}\tau )\bigg]
	\end{equation}
	where the index $j$ labels the four eigenvalues corresponding to two left and right movers each. The subscript $j$ on $f$ which is the eigenvalue index should not be confused with the subscript $k$ on quantities like $\rho,\;v,\; \psi$ which is the species index. We choose $j=1,2$ denoting both right movers with speed $+\sqrt{\frac{A\pm B}{2}}$. In order to compare with the coupled NLS, $\delta\rho$ and $\delta v$ need to be calculated from \eqref{relation1}. For the fastest mover, $j=1$ eigenvalue $\lambda_1$ is chosen.

	\begin{align}
		\begin{pmatrix}
		\delta\rho_1\vspace{8pt}\\
		\delta\rho_2\vspace{8pt}\\
		\delta v_1\vspace{8pt}\\
		\delta v_2
		\end{pmatrix}=
		f_1(\xi,\tau) \begin{pmatrix}
		\dfrac{C+B}{2h\lambda_1} \vspace{8pt}\\
		\dfrac{\rho_{02}}{\lambda_1}\vspace{8pt}\\
		\dfrac{C+B}{2h\rho_{01}}\vspace{8pt}\\
		1
		\end{pmatrix}
	\end{align}

	We can now obtain $\rho(x,t)$ and $v(x,t)$ from \eqref{relation2}, given, $\xi  = \epsilon(x-\lambda_j t)$ and $\tau = \epsilon^3 t$.  

	\begin{subequations}
		\begin{align}
			\rho_1(x,t)&=\rho_{01}+\epsilon^2\frac{C+B}{2h\lambda_1}f_1(\xi,\tau)\\
			\rho_2(x,t)&=\rho_{02}+\epsilon^2\frac{\rho_{02}}{\lambda_1}f_1(\xi,\tau)\\
			v_1(x,t)&=\epsilon^2\frac{C+B}{2h\rho_{01}}f_1(\xi,\tau)\\
			v_2(x,t)&=\epsilon^2f_1(\xi,\tau)
		\end{align} 
	\end{subequations}

	Subsequently, $\psi(x,t)$ can be obtained by using the transformation \eqref{eqn:madelung}. This requires the calculating the integral $\int_{0}^{x}\text{sech}^{2}(ax',t)dx'=\frac{1}{a}\tanh(ax,t)$. By repeating this exercise for the next larger eigenvalue $\lambda_{j=2}$, we thus obtain the initial profiles for both the components as in \eqref{eqn:initfast},\eqref{eqn:initslow}.

	We simulate the dynamics of both the right chiral sectors having speeds $\lambda_1$ and $\lambda_2$. The initial conditions in the NLS language corresponding to the two eigenvalues $\lambda_1,\lambda_2$ are
	\begin{widetext}
		\begin{subequations}\label{eqn:initfast}
		\begin{align}
		\psi^{\lambda_1}_1(x,0)&=\sqrt{\rho_{01}+\epsilon^2\frac{(C+B)}{4h\lambda_1}f_1(  \epsilon x,0)}\;\exp\bigg[\frac{i\epsilon}{\hbar} \frac{6A_1(C+B)\sqrt{\mathcal{V}}}{2hB_1\rho_{01}}\tanh(\frac{\sqrt{\mathcal{V}}}{2} \epsilon x)\bigg]
		\\
		\psi^{\lambda_1}_2(x,0)&=\sqrt{\rho_{02}+\epsilon^2\frac{\rho_{02}}{\lambda_1}f_1( \epsilon x,0)}\;\exp\bigg[\frac{i\epsilon}{\hbar}\frac{6A_1\sqrt{\mathcal{V}}}{B_1}\tanh(\frac{\sqrt{\mathcal{V}}}{2}  \epsilon x)\bigg]
		\end{align}
		\end{subequations}
		\begin{subequations}\label{eqn:initslow}
		\begin{align}
			\psi^{\lambda_2}_1(x,0)&=\sqrt{\rho_{01}+\epsilon^2\frac{(C-B)}{2h\lambda_2}f_2( \epsilon x,0)}\;\exp\bigg[\frac{i\epsilon}{\hbar}\frac{6A_2(C-B)\sqrt{\mathcal{V}}}{2hB_2\rho_{01}}\tanh(\frac{\sqrt{\mathcal{V}}}{2} \epsilon x)\bigg]
			\\
			\psi^{\lambda_2}_2(x,0)&=\sqrt{\rho_{02}+\epsilon^2\frac{\rho_{02}}{\lambda_2}f_2( \epsilon x,0)}\;\exp\bigg[\frac{i\epsilon}{\hbar}\frac{6A_2\sqrt{\mathcal{V}}}{B_2}\tanh(\frac{\sqrt{\mathcal{V}}}{2}  \epsilon x)\bigg]
		\end{align}
		\end{subequations}
	\end{widetext}
	For simulating the coupled NLS, we use Classical Explicit Method as the time stepping method as outlined in Ref.~\onlinecite{tahaablowitz}. Keeping in mind the conditions on coupling constants, we choose the following values for the parameters: $g_1=1, \; g_2=1,\; h=0.5,\; \epsilon=0.2,\; \rho_{01}=1,\; \rho_{02}=0.1$. Thus, $A=1.1$, $B=0.954$ and $C=0.9$. The system size is $L=300$ and spacial axis runs from $x=-150 \text{ to } 150$ which has been discretized into $n=12000$ steps. Thus, $dx= 0.025$ and $dt= 7.8125\times10^{-5}$. The speed of the solitary wave is chosen as $\mathcal{V}=2.5$.
	
	Now that we have the correct initial conditions for both NLS \eqref{eqn:initfast} and KdV \eqref{eqn:initkdv}, we let them evolve in time. We study the time evolution of $|\psi_k(x,t)|^2$. At this stage, the coupled NLS and KdV profiles cannot be compared because they are not the same physical quantities. Since, the KdV problem only involves $f_j$'s, while the NLS profile $\psi_k^{\lambda_j}$ is a combination of $f_j$'s multiplied by appropriate coefficients and added to a background, for a meaningful comparison, the KdV variables need to be rescaled via a suitable transformation. The appropriate physical quantities are the density fields and can be obtained by using the transformation \eqref{rpert}. We also take care that the two profiles are in the same frame of reference. We choose the lab frame $(x,t)$ for both the profiles. Since, $f_j(\xi,\tau)$ is a function of $(\xi-A_j\mathcal{V}\tau)$, using the inverse scaling relations we obtain the speeds in the lab frame. 
	\begin{equation}
		\xi -A_j\mathcal{V}\tau = \epsilon x- \epsilon t (\lambda_j+A_j\mathcal{V}\epsilon^2)=\epsilon(x-\varLambda_j t)
	\end{equation}
	Thus, the speed of sound in the lab frame is $\varLambda_j=\lambda_j+A_j\mathcal{V}\epsilon^2$. For the chosen coupling constants, the parameters are $A_1=-0.123$, $B_1=1.372$, $A_2=-0.462$, $B_2=0.728$, $\lambda_1=1.013$, $\lambda_2=0.270$, $\varLambda_1=1.001$, and $\varLambda_2=0.224$. Note that opposite signs of $A_j$ and $B_j$ for both $j=1,2$ ensure an overall negative sign on $f_j(x,t)$ \eqref{eqn:initkdv}. Additionally, since the coefficient $C-B=-0.054<0$,  $|\psi_1^{\lambda_2}|^2$ profile is expected to be a bump as opposed to  the other profiles being a dip. For completeness, we explicitly write out the density fields for both eigenvalues $j=1,2$ for both components $k=1,2$, given, $\xi  = \epsilon(x-\lambda_j t)$ and $\tau = \epsilon^3 t$, 
	\begin{subequations}\label{eqn:rhophys}
		\begin{align}
		\rho^{\lambda_1}_1(x,t)&=\rho_{01}+\epsilon^2\frac{(C+B)}{4h\lambda_1}f_1(\xi,\tau)\label{eqn:rhophysA}\\
		\rho^{\lambda_1}_2(x,t)&=\rho_{02}+\epsilon^2\frac{\rho_{02}}{\lambda_1}f_1(\xi,\tau)\label{eqn:rhophysB}\\
		\rho^{\lambda_2}_1x,t)&=\rho_{01}+\epsilon^2\frac{(C-B)}{2h\lambda_2}f_2(\xi,\tau)\label{eqn:rhophysC}\\
		\rho^{\lambda_2}_2(x,t)&=\rho_{02}+\epsilon^2\frac{\rho_{02}}{\lambda_2}f_2(\xi,\tau)\label{eqn:rhophysD}
		\end{align}
	\end{subequations}
	We plot four different time snapshots of the evolution of $|\psi_k(x)|^2$ through the coupled NLS equation \eqref{nls} along with the evolution of $\rho_k(x)$ through the KdV equation \eqref{eq:kdvN2} in Fig. (\ref{fig:a}) for the largest eigenvalue ($\lambda_1$) and in Fig. (\ref{fig:b}) for the second-largest eigenvalue  ($\lambda_2$) . It is surprising to note that they have quantitative agreement for significant times. Thus, it turns out that if we just evolve the NLS problem with the initial conditions \eqref{eqn:initfast},\eqref{eqn:initslow} without the knowledge of any scaling or transformations applied, the evolution has a significant match with the independent KdV evolution of \eqref{eqn:rhophys}. Consequently, we see a strong correspondence between the two equations, namely, coupled NLS and the KdV equation. 

	\begin{figure}
		\centering
		\includegraphics[width=.9\linewidth]{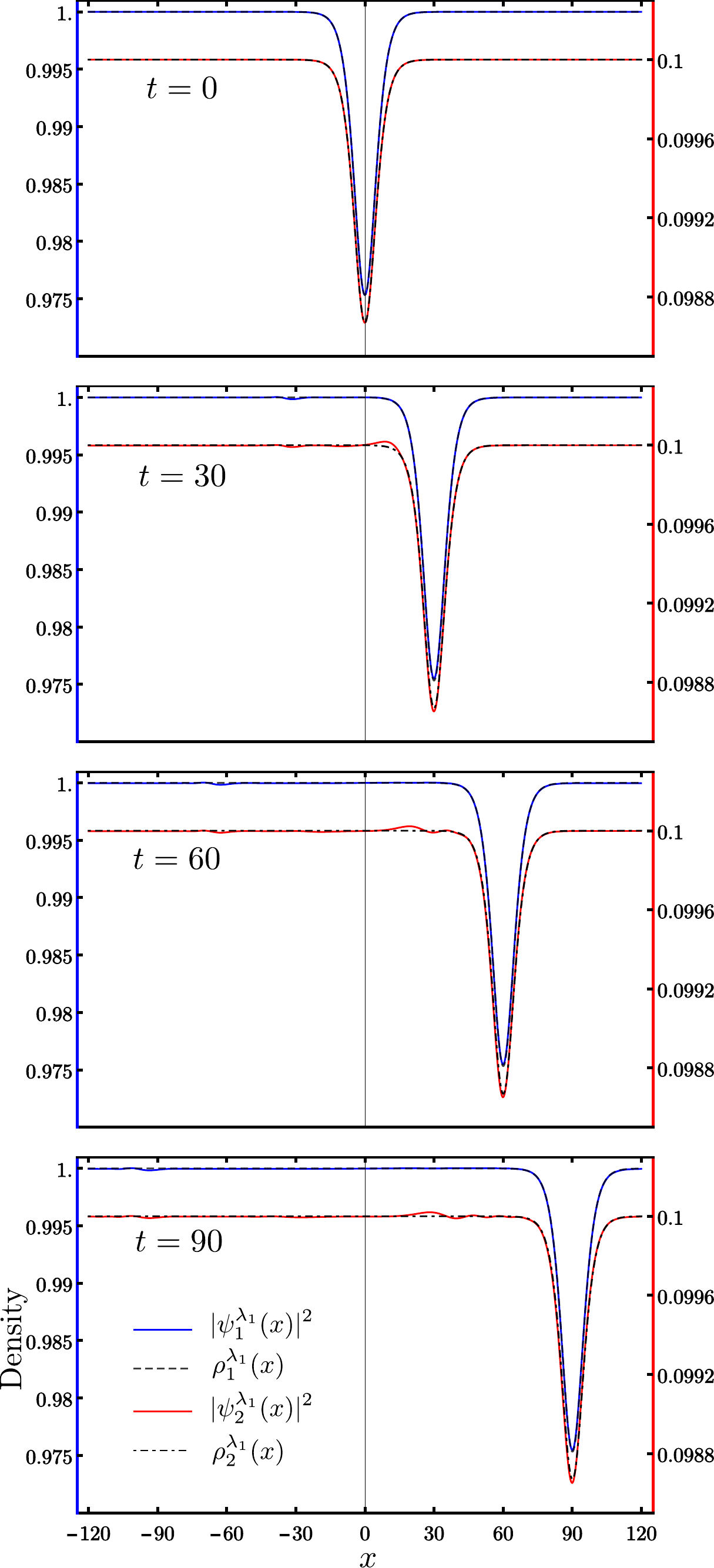}
		\caption{\vv{Evolution of the densities $|\psi_k^{\lambda_1}(x)|^2$ with time for $k=1,2$ species starting from the initial profile \eqref{eqn:initfast}. The blue (species $1$) and the red (species $2$) plots are evolve under the binary coupled NLS equation \eqref{nls}. The left $y-$axis (blue) is the scale for species $1$ while the right $y-$axis (red) for species $2$ (since the background densities differ significantly). The speed of sound (eigenvalue) is $\lambda_1=+\sqrt{(A+B)/2}$ and in the lab frame has a value $\varLambda_1=1.001$. Independently, the analytical density plot $\rho_k(x,t)$ of the KdV soliton (\ref{eqn:rhophysA}-\ref{eqn:rhophysB}) is also shown for comparison. Evidently  the four well separated time snapshots indicate both species exhibit a remarkable match to the analytical 
		density plots (dashed and dot-dashed for species 1,2 respectively) with emission of some small amount of radiation in both directions.} }
		\label{fig:a}
	\end{figure}
	\begin{figure}
		\centering
		\includegraphics[width=0.9\linewidth]{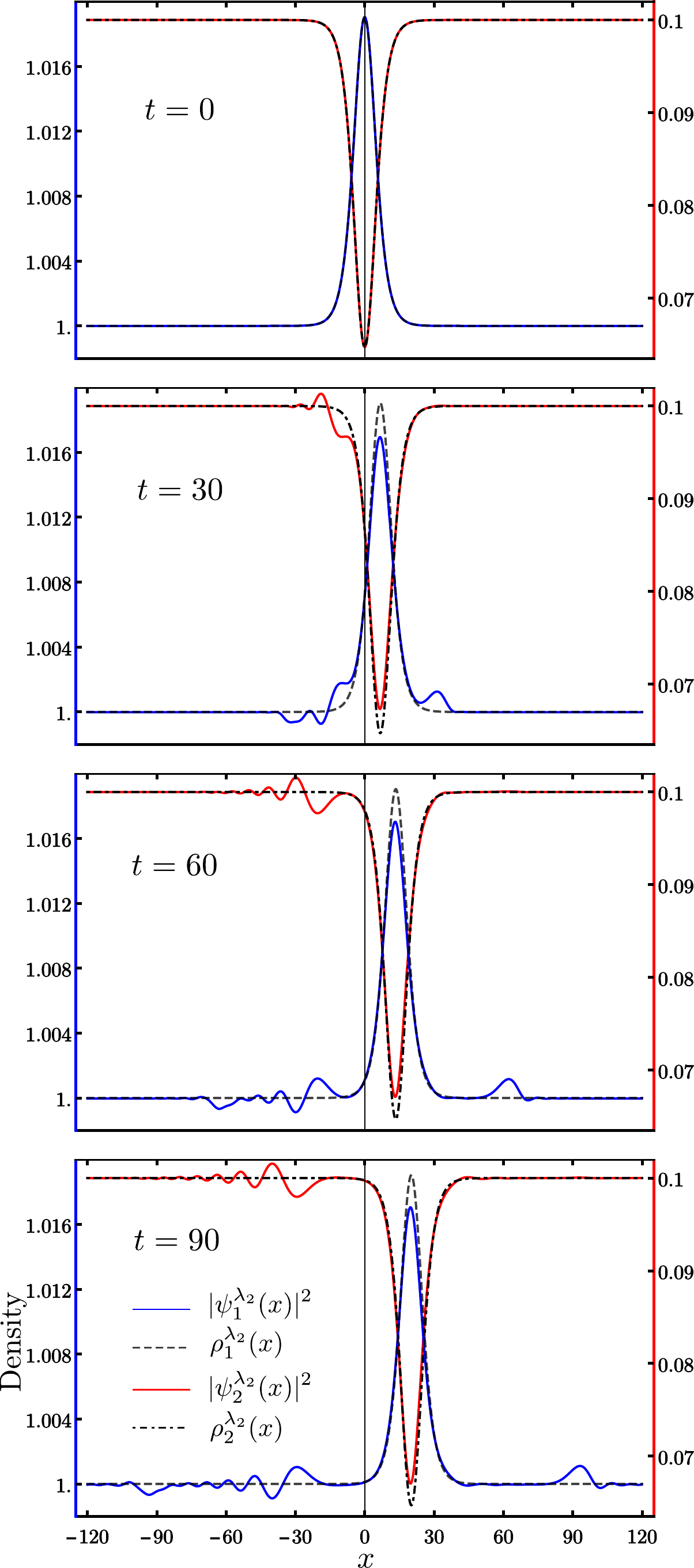}
		\caption{\vv{Evolution of the density fields $|\psi_k^{\lambda_2}(x)|^2$ in time for $k=1,2$ species starting from the initial profile \eqref{eqn:initslow} under the dynamics of the coupled NLS equation \eqref{nls}. The largest speed of sound  (eigenvalue) $\lambda_1$ is shown in Fig. (\ref{fig:a}). Here the secondand largest speed $\lambda_2=+\sqrt{(A-B)/2}$ is considered with a lab-frame speed of $\varLambda_2=0.224$. In addition, the analytical density $\rho_k(x,t)$ of the KdV soliton (\ref{eqn:rhophysC}-\ref{eqn:rhophysD}) is shown for comparision. Here too we find that both species exhibit a remarkable match with the analytical expressions (dashed and dot-dashed respectively) and emission of some small amount of radiation in both directions. A very interesting aspect about this eigenvalue ($\lambda_2$) is that it results in species 1 having a bump and species 2 having a dip.} }
		\label{fig:b}
	\end{figure}

	\section{\label{sec:VI} Conclusions}

	In this paper, we have analyzed the linear and nonlinear problem for the multi-component NLS which is a physically relevant system spanning a broad range of fields. We have systematically studied the qualitative long time dynamics of non-equilibrium profiles. We started with writing a hydrodynamic form. In the linearized regime, we stated and proved a set of theorems. We obtained necessary and sufficient conditions for real speeds of sound that depend solely on the coupling matrix. For the nonlinear problem,  using the key mathematical concept of the Fredholm alternative, we show that the coefficients of the KdV dynamics are given in terms of the eigenvectors of the linearised problem.  We also discuss numerical protocols to compute the eigenvectors of the coupling matrix individually that also provides us with information on how eigenvalues (sound speeds) and eigenvectors (KdV coefficients) change as the cross-component coupling coefficient varies. This is of high experimental relevance given the tunability of coupling constants. We show compelling evidence of agreement between KdV and multi-component NLS in the nonlinear dynamics using soliton profiles as a platform for comparison This kind of effective mapping shines light on the complex non-equilibrium dynamics of interacting multi-component coupled systems. 
 
The present manuscript investigated the qualitative dynamics of small amplitude perturbations of a trivial state when the speeds of sound (namely the eigenvalues $\mathcal{A}$) are distinct The case of repeated eigenvalues leading to coupled KdV will be investigated in future works. The future outlook also includes generalizations of the coupling matrix $\alpha$. This is important given various physical systems where coupling can vary spatially. Understanding the role of external potential from a rigorous perspective remains unexplored. Although this work is restricted to Hamiltonian systems, it can be extended to open systems which are connected to reservoirs \cite{urb} and much remains unexplored in that avenue of driven-dissipative systems. 

	\section*{Acknowledgments}
	
We thank Bernard Deconinck, Chiara D'Errico, Nicolas Pavloff, Urbashi Satpathi and Raghavendra Nimiwal  for useful discussions. MK gratefully acknowledges the Ramanujan Fellowship SB/S2/RJN-114/2016 from the Science and Engineering Research Board (SERB), Department of Science and Technology, Government of India and support the Early Career Research Award, ECR/2018/002085  from the Science and Engineering Research Board (SERB), Department of Science and Technology, Government of India. MK would like to acknowledge support from the project 6004-1 of the Indo-French Centre for the Promotion of Advanced Research (IFCPAR). SS acknowledges ICTS for support and hospitality during the S. N. Bhatt Memorial Excellence Fellowship Program 2017. 

\appendix

\section{Proof of Theorem \ref{thm:alpha:PosDef}} \label{p-thm:alpha:PosDef} 

 Suppose $\alpha$ is positive definite. From Sylvester's criterion, determinants of all leading principal minors of a positive definite matrix are positive. Then under the supposed ordering, the determinant of the $2\times 2$ leading principal minor, \emph{i.e.} the $2\times 2$ matrix in the top left corner, $g_1g_2-h^2$ is positive. Hence $h<\sqrt{g_1g_2}$.

	\section{Proof of Theorem \ref{thm:CharPoly}} \label{app:proof:charpoly}

    \noindent We first introduce a definition for the symmetric products of $\gamma_i$. 
    {\definition{
        
    We denote the set of $\{\gamma_i\}$, obtained by eliminating $\gamma_k$ for some $k$, by $\{\gamma_i\} - \gamma_k$. Similarly, the set obtained adding an element $\gamma_k$ is denoted by $\{\gamma_i\}+\gamma_k$. Suppose $\gamma_j=\gamma_k$ for some $j,k$ for a list of $\{\gamma_i\}$. Let $\beta_i$ be the distinct $\gamma_i$. Then $\{\gamma_i\} = \{\beta_i\} + \gamma_k$, where $\beta_i$ are all distinct. Consequently we describe the replacement of $\gamma_k$ with $\gamma_j$ in a list by $\{\gamma_i\} - \gamma_j + \gamma_k$.
    }} 
    Keeping in mind that \begin{equation}
    	\Sym{\gamma_i}{0} = 1, \quad \Sym{\gamma_i}{k} = 0,\ k>N, 
    \end{equation} we have the following consequence of these definitions.

    {\theorem{ \label{thm:sym_add}
        \begin{equation}
        	\SymPlus{\gamma_i}{+\gamma_k}{m} =  \gamma_k\Sym{\gamma_i}{m-1} + \Sym{\gamma_i}{m}.
        \end{equation}
        In particular, suppose one of $\gamma_i$ equals $1$. Then setting $\{\gamma_i\}=\{\beta_i\} + 1$ we have
        \begin{equation}
        	\SymPlus{\beta_i}{+1}{m} = \Sym{\beta_i}{m} + \Sym{\beta_i}{m-1}.
        \end{equation}
    }}

    \noindent Theorem \ref{thm:CharPoly} is proved using the principle of induction. Consider the case when $\alpha,\rho$ are $2\times 2$ matrices. Then
    \begin{align}\nonumber
    \mbox{det}(\mu - \rho\alpha) &= \mbox{det}\left(\begin{array}{cc} \mu- \rho_1g_1 & -\rho_1 h, \\ -\rho_2 h & \mu -  \rho_2g_2  \end{array} \right) \\
    &=  \rho_1\rho_2 h^2 \left( \frac{\mu-\rho_1g_1}{\rho_1 h} \ \frac{\mu-\rho_2g_2}{\rho_2 h}- 1\right) \nonumber\\
    & = \rho_1\rho_2h^2 \ \mathcal P(\mu),
    \end{align}
    and so the theorem is true for $N=2$. Furthermore, note that since $\alpha\rho = \rho^{-1}\rho\alpha\rho$, then $\alpha\rho$ and $\rho\alpha$ have the same eigenvalues. Let us now suppose the theorem holds for matrices of some size $n$. We denote the relevant matrices by $\rho^{(n)}$, $\alpha^{(n)}$ so that $\rho^{(n)}\alpha^{(n)}-\mu$ is given by 
    \begin{equation}
    	(\rho^{(n)}\alpha^{(n)}-\mu)_{ij}=\left\{\begin{array}{ll} \rho_{0n-i} g_{n-1} - \mu, & i=j, \\ \rho_{0n-i}h, & i\neq j \end{array} \right. 
    \end{equation}
    where $\mu$ is an eigenvalue of $\rho^{(n)}\alpha^{(n)}$. The characteristic polynomial is
    \begin{equation}
    	 \mbox{det}(\rho^{(n)}\alpha^{(n)} - \mu) = 0 \Rightarrow \prod_{i=1}^n \rho_{0i}\: h^n\: \mbox{det}(X_n) = 0 
    \end{equation}
    where \[ (X_n)_{ij} = \left\{\begin{array}{ll}\gamma_{n+1-i} , & i=j, \\ 1, & i\neq j, \end{array} \right. \mbox{ and} \quad \gamma_{i} =  \dfrac{\rho_{0i} g_{i} - \mu}{\rho_{0i} h }.\]
    Since the theorem holds for the $n-$th order matrix, we have
    \begin{align} 
    \det(X_n) &= \Sym{\gamma_i}{n} \nonumber \\
    &\quad + \sum_{k=2}^{n}(-1)^{k-1}(k-1) \Sym{\gamma_i}{n-k} = 0.
    \end{align}
    The matrix $X_{n+1}$ is obtained from $X_n$ by the following relation
    \begin{equation}
    	 X_{n+1} = \left(\begin{array}{cc} \gamma_{n+1} & \vec{1}^{\ T}_n \\ & \\ \vec 1_n & X_n \end{array} \right)
    \end{equation}
    where $\vec 1_n$ is a column vector of length $n$ consisting of only ones. The characteristic polynomial then is obtained by computing $\det(X_{n+1})$. This determinant is obtained by a linear combination of the determinants of its co-factor matrices. The co-factor matrices of $X_{n+1}$ are either $X_{n}$ or the matrix obtained by replacing the relevant column of $X_n$ by $\vec 1$. The determinants of those co-factor matrices, obtained by replacing a column of $X_n$ by $\vec 1$, are equal to (up to sign) the determinants of matrices $X_{n}^{(j)}$ where 
    \begin{equation}
    	 (X_{n}^{(j)})_{ik} = \left\{\begin{array}{ll} 1, & i=k=j, \\ (X_n)_{ik}, & \mbox{otherwise}. \end{array}\right.  
    \end{equation}Taking into account the signs, we  have
    \begin{align} 
    \det(X_{n+1}) =  \gamma_{n+1} \det(X_n) - \sum_{j=1}^n \det(X_n^{(j)}). \label{eqn:induction}
    \end{align}
    We consider each term on the right-hand side of the above equation individually. By definition $\det(X_n^{(j)})$
    \begin{equation}
    	\begin{aligned} 
    	&= \SymPlus{\gamma_i}{-\gamma_j+1}{n}\\
    	& + \sum_{k=2}^n (-1)^{k-1}(k-1)\SymPlus{\gamma_i}{-\gamma_j+1}{n-k},\\
    	&= \SymPlus{\gamma_i}{-\gamma_j}{n} + \SymPlus{\gamma_i}{-\gamma_j}{n-1} \\
    	&+ \sum_{k=2}^{n-1} (-1)^{k-1}(k-1)\SymPlus{\gamma_i}{-\gamma_j}{n-k} \\
    	& + \sum_{k=2}^{n-1} (-1)^{k-1}(k-1)\SymPlus{\gamma_i}{-\gamma_j}{n-k-1} \\
    	&+ (-1)^{n-1}(n-1),
    	\end{aligned}
    \end{equation}
    where we have used theorem \ref{thm:sym_add}. It is straightforward to show that if a set $\{\beta_i\}$ contains $n$ elements, then for $m<n$
    \begin{align} \nonumber
    \sum_{j=1}^n \SymPlus{\beta_i}{-\beta_j}{m} &= \sum_j\left[ \Sym{\beta_i}{m}- \beta_j \SymPlus{\beta_i}{-\beta_j}{m-1}\right],\\
     &= (n-m)\Sym{\beta_i}{m}.
    \end{align}
    Using this relation we have
    \begin{equation}
    	\begin{aligned}
    	\sum_{j=1}^n \det(X_n^{(j)}) &= \Sym{\gamma_i}{n-1}  \\
    	&+ \sum_{k=2}^{n-1}(-1)^{k-1}k(k-1)\Sym{\gamma_i}{n-k} \\
    	&+ \sum_{k=2}^{n-2}(-1)^{k-1}(k-1)(k+1)\Sym{\gamma_i}{n-k-1}\\
    	& +(-1)^{n}(n-2)n - (-1)^{n}(n-1)n\\
    	&=  - \sum_{k=1}^{n}(-1)^{k}k\ \Sym{\gamma_i}{n-k}.
    	\end{aligned}
    \end{equation}
    On the other hand,$\gamma_{n+1}\det{(X_{n})}$
    \begin{equation}
    	\begin{aligned}
    	&= \gamma_{n+1}\Sym{\gamma_i}{n} + \sum_{k=2}^{n}(-1)^{k-1}(k-1)\gamma_{n+1}\Sym{\gamma_i}{n-k},\\
    	&= \SymPlus{\gamma_i}{+\gamma_{n+1}}{n+1} \\
    	&\quad + \sum_{k=2}^n(-1)^{k-1}(k-1)\SymPlus{\gamma_i}{+\gamma_{n+1}}{n+1-k} \\
    	&\quad - \sum_{k=2}^n(-1)^{k-1}(k-1)\Sym{\gamma_i}{n+1-k}\\
    	&= \SymPlus{\gamma_i}{+\gamma_{n+1}}{n+1} \\
    	&\quad + \sum_{k=2}^{n+1}(-1)^{k-1}(k-1)\SymPlus{\gamma_i}{+\gamma_{n+1}}{n+1-k}\\
    	&\quad  - \sum_{k=1}^n(-1)^k k \ \Sym{\gamma_i}{n-k}
    	\end{aligned}
    \end{equation}

    Combining the expressions for either of the right-hand side terms of (\ref{eqn:induction}) we have 
    \begin{align} \nonumber
    \det{(X_{n+1})} &= \SymPlus{\gamma_i}{+\gamma_{n+1}}{n+1} \\
    &\quad + \sum_{k=2}^{n+1}(-1)^{k-1}(k-1)\SymPlus{\gamma_i}{+\gamma_{n+1}}{n+1-k}, 
    \end{align}
    which proves the theorem for $n+1$ and the statement of the Theorem \ref{thm:CharPoly} follows.

	\section{Proof of Theorem \ref{thm:A:multiple}} \label{app:proof:multiple}
    \noindent Let $\mathcal P(\lambda^2)$ denote the characteristic polynomial of $\alpha\rho$. We first claim that repeated eigenvalues can only occur when $N>2$. This is easy to see since the characteristic polynomial for $N=2$ is readily computed as 
    \begin{equation}
    	 \mathcal P(\lambda^2) = \left(\frac{\rho_{01} g_1-\lambda^2}{\rho_{01} h} \right)\left( \frac{\rho_{02} g_2-\lambda^2}{\rho_{02} h} \right) - 1,
    \end{equation}
    which has roots 
    \begin{align}
    \lambda^2 &= \frac{(\rho_{01}g_1+\rho_{02}g_2) \pm \sqrt{(\rho_{01}g_1-\rho_{02}g_2)^2+4\rho_{02}\rho_{01}h^2} }{2\rho_{01}\rho_{02}h^2}.
    \end{align}
    Since the discriminant is positive for all real values of $\rho_{0i},g_i,h$, there are no repeated roots when $N=2$. 
    \subsection{Proof of sufficiency}

    \noindent We now proceed to the general case. We consider first $m$ repeated pairs of $(\rho_{0i}g_i,\rho_{0i})$ and prove this is a sufficient condition to guarantee a repeated eigenvalue. Let $(\rho^*_0,g^*)$ denote the common value of $m$ repeated pairs of $(\rho_{0i},g_i)$. Since the total number of eigenvalue pairs is $N$, there are $N-m$ not-necessarily-repeated pairs. Define 
    \begin{equation}
    	 \gamma = \frac{\rho_0^* g^* - \lambda^2}{\rho^*_0 h},\quad \gamma_i = \frac{\rho_{0i} g_i - \lambda^2}{\rho_{0i} h},
    \end{equation}
    where the $\gamma_i$ are shorthand for the not necessarily repeated pairs. The characteristic polynomial is
    \begin{align}\nonumber
    &\SymPlus{\gamma_i}{+m\gamma}{N} + \\
    \quad &\sum_{k=2}^N (-1)^{k-1}(k-1)\SymPlus{\gamma_i}{+m\gamma}{N-k} = 0.
    \end{align}
    Here $\{\gamma_i\}+m\gamma$ in the argument to the symmetric polynomial indicates $m$ repetitions of $\gamma$ in addition to the list of $\gamma_i$. We note that 
    \begin{equation}
    	\SymPlus{\gamma_i}{+m\gamma}{l} = \sum_{p=0}^l \Sym{\gamma_i}{p}\Sym{m \gamma }{l-p}.
    \end{equation}
    Also notice that $\SymPlus{\gamma_i}{+m\gamma}{N} = \gamma^m \Sym{\gamma_i}{N-m}$ which also follows from the identity above when $l=N$ and recalling $\Sym{\beta_i}{l}=0$ when $l$ is larger than the number of elements in the list $\{\beta_i\}$. Substituting this identity into the expression for the characteristic polynomial we have
    \begin{equation}
    	\begin{aligned}
    	0  &= \SymPlus{\gamma_i}{+m\gamma}{N}\\
    	&\: + \sum_{k=2}^N (-1)^{k-1}(k-1)\SymPlus{\gamma_i}{+m\gamma}{N-k},\\
    	&= \gamma^m \Sym{\gamma_i}{N-m}  \\
    	&\:  + \sum_{k=2}^N (-1)^{k-1}(k-1)\sum_{p=0}^{N-k}\ \Sym{\gamma_i}{p}\ \Sym{m \gamma }{N-k-p},\\
    	&= \gamma^m \Sym{\gamma_i}{N-m} \\
    	&\: + \sum_{p=0}^{N-2}\ \sum_{k=2}^{N-p}(-1)^{k-1}(k-1)\ \Sym{\gamma_i}{p}\ \Sym{m \gamma }{N-k-p},\\
    	&= \gamma^m \Sym{\gamma_i}{N-m} \\
    	&\: + \sum_{p=0}^{N-m}\ \sum_{k=2}^{N-p}(-1)^{k-1}(k-1)\ \Sym{\gamma_i}{p}\ \Sym{m \gamma}{N-k-p},\\
    	&= \Sym{\gamma_i}{N-m}[ \gamma^m + \sum_{k=2}^{m}(-1)^{k-1}(k-1)\Sym{m \gamma }{m-k} ] \\
    	&\: + \sum_{p=0}^{N-m-1} \Sym{\gamma_i}{p}\  \sum_{k=2}^{N-p}(-1)^{k-1}(k-1)\ \Sym{m \gamma }{N-k-p}
    	\end{aligned}
    \end{equation}

    Note that $\Sym{m\gamma}{m-k}$ represents products of $m-k$ $\gamma$'s. Of course there are $^{m}C_{m-k}$ ways to choose these products and thus we have
    \begin{equation}
    	\begin{aligned}
    	\gamma^ m +& \sum_{k=2}^{m}(-1)^{k-1}(k-1)\Sym{m\gamma}{m-k}\\ 
    	&= \gamma^m + \sum_{k=2}^m (-1)^{k-1}(k-1) \frac{\gamma^{m-k} m!}{(m-k)!k!} \\
    	&= \sum_{k=0}^{m}(-1)^{k-1}(k-1)\frac{\gamma^{m-k}m!}{(m-k)!k!}\\
    	&= m(\gamma-1)^{m-1} + (\gamma-1)^m .
    	\end{aligned}
    \end{equation}
    Substituting the above in to the expression for the characteristic polynomial we have
    \begin{equation}
    	\begin{aligned}
    	&\Sym{\gamma_i}{N-m}\left[ m(\gamma-1)^{m-1} + (\gamma-1)^m \right]\\
    	&+ \sum_{p=0}^{N-m-1} \Sym{\gamma_i}{p}
    	\sum_{k=2}^{N-p}(-1)^{k-1}(k-1) \Sym{m\gamma}{N-k-p} = 0.
    	\end{aligned}
    \end{equation}
    Evidently, the above expression is true for $m<N$. Indeed if $m=N$ we have
    \begin{align}\nonumber
  	\mathcal P(\lambda^2) &= \Sym{\gamma_i}{N-m}\left[ m(\gamma-1)^{m-1} + (\gamma-1)^m \right]\\
    & = (\gamma-1)^{N-1}(N-1 + \gamma ), 
    \end{align}
    or in other words
    \begin{equation}
    	 \mathcal P(\lambda^2) = \left(\frac{\rho^*_0 g^* -\lambda^2}{\rho^*_0 h} - 1\right)^{N-1}\left(N - 1 + \frac{\rho^*_0 g^* -\lambda^2}{\rho^*_0 h} \right). 
    \end{equation}
    Moving ahead with the case $m<N$, we have
    \begin{equation}
    	\begin{aligned}
    	&\Sym{\gamma_i}{N-m}(\gamma-1)^{m-1}(m-1 + \gamma)\\
    	&+\sum_{p=0}^{N-m-1}\! \Sym{\gamma_i}{p}
    	\sum_{k=2}^{N-p}(-1)^{k-1}(k-1)\Sym{m\gamma}{N-k-p} = 0
    	\end{aligned}
    \end{equation}
    The second term on the left-hand side may be simplified as follows
    \begin{equation}
    	\begin{aligned}
    	&\sum_{k=2}^{N-p}(-1)^{k-1}(k-1) \Sym{m\gamma}{N-k-p}\\
    	&= \sum_{k=N-p-m}^{N-p}\hspace{-5pt}(-1)^{k-1}(k-1)\Sym{m\gamma}{N-k-p},\\
    	&= \sum_{l=0}^m (-1)^{N-p-m+l-1}(N-p-m+l-1)\Sym{m\gamma}{m-l},\\
    	&=(-1)^{N-p-m} \sum_{l=0}^m (-1)^{l-1}(N-p-m-1+l) \frac{\gamma^{m-l}\ m!}{(m-l)!\ l!},\\
    	&= (-1)^{N-p-m} \left[ m(\gamma-1)^{m-1} - (N-p-m-1)(\gamma-1)^{m}\right],\\
    	&= (-1)^{N-p-m}(\gamma-1)^{m-1}(m+(1-N+p+m)(\gamma-1)).
    	\end{aligned}
    \end{equation}
    The first equality is true since $\Sym{m\gamma}{N-k-p}=0$ unless $N-k-p\leq m$. To replace the lower limit of the $k-$sum we also need to assure $N-p-m\geq 2$ which implies $p\leq N-m-2$. Thus the only possible exception is when $p=N-m-1$, \emph{i.e.} the upper limit of the $p-$sum. However, it is easy to see that this term has no contribution for $k=0,1$ since $\Sym{m\gamma}{m+1-k}=0$ and $k-1=0$ when $k=0,1$ respectively. Finally we obtain the following expression for the characteristic polynomial
    \begin{align} \nonumber
    &\mathcal P(\lambda^2) =  \left(\frac{\rho^*_0 g^* -\lambda^2}{\rho^*_0 h} - 1\right)^{m-1} \times \\ \nonumber
    &\left[ \sum_{p=0}^{N-m} \Sym{\frac{\rho_{0i} g_i -\lambda^2}{\rho_{0i} h}}{p} \times \right. \\
    &\left. (-1)^{N-p-m}\left(m+(1-N+p+m)\left(\frac{\rho^*_0 g -\lambda^2}{\rho^*_0 h} - 1\right)\right)\right] .
    \end{align}
    \subsection{Proof of necessity}
    It is also necessary that at least three pairs of $(\rho_{0i}g_i,\rho_{0i})$ be equal for the characteristic polynomial to be permanently degenerate. To show this we prove the contrapositive, \emph{i.e.} we show that if only $m=1,2$ of the $(\rho_{0i} g_i,\rho_{0i})$ pairs are equal, then the polynomial is not permanently degenerate. Consider first the case $m=1$ \emph{i.e.} when none of the $(\rho_{0i} g_i,\rho_{0i})$ are equal. Then a standard implicit function theorem argument applied to 
    \begin{align} \nonumber
    \mathcal Q(\mu) \coloneqq &\mathcal P(\mu)h^n = \Sym{\frac{\rho_{0i} g_i - \mu}{\rho_{0i}}}{N}\\
     +& \sum_{k=2}^{N} (-1)^{k-1}(k-1)\ h^k \Sym{\frac{\rho_{0i} g_i - \mu}{\rho_{0i}}}{N-k},
    \end{align}
    using the fact that (i) $\mathcal Q(\mu)$ has distinct zeros $\rho_{0i} g_i$ when $h=0$ and, (ii) $\partial \mathcal Q/\partial \mu$ evaluated at $h=0$ is non-zero (due to the distinct values of $\rho_{0i},g_i$), we have an open neighborhood of $h=0$ where there are $N$ distinct roots to the polynomial $\mathcal Q$ and so the polynomial cannot be permanently degenerate. We now consider the case $m=2$. Suppose only two of the $(\rho_{0i} g_i,\rho_{0i})$ are equal. Let the common value be $(\rho^*_0 g^*,\rho^*_0)$. By assumption this common value is distinct from all remaining $N-2$ $(\rho_{0i} g_i,\rho_{0i})$ pairs. Factoring as we did previously, but for the case $m=2$, we have \vspace{-10pt}
    \begin{align} 
    &\mathcal Q(\mu) =  \left(\frac{\rho^*_0 g^* -\mu}{\rho^*_0 h} - 1\right) \times \nonumber \\
    &\left[ \sum_{p=0}^{N-2} \Sym{\frac{\rho_{0i} g_i -\mu}{\rho_{0i} h}}{p} (-1)^{N-p-2}\: \times \nonumber\right. \\
    & \left.\left(2+(1-N+p+2)\left(\frac{\rho^*_0 g^* -\mu}{\rho^*_0 h} - 1\right)\right)\right] .
    \end{align}
    This polynomial has a root $\mu=\rho^*_0(g^*-h)$ and the roots of 
    \begin{align} \nonumber
    &\sum_{p=0}^{N-2} \Sym{\frac{\rho_{0i} g_i -\mu}{\rho_{0i} h}}{p}\times \\ 
    &(-1)^{N-p-2} \left(2+(1-N+p+2)\left(\frac{\rho^*_0 g -\mu}{\rho_0^* h} - 1\right)\right)=0.
    \end{align}
    Applying the implicit function theorem to the above polynomial we have that in an open neighborhood of $h=0$, the above polynomial has distinct roots. The only remaining possibility is that $\rho^*_0(g^*-h)$ itself is a root of this polynomial. But this leads to the following expression valid for all suitable $h$.
    \begin{align} 
    \sum_{p=0}^{N-2} \Sym{\frac{\rho_{0i} g_i -\rho^*_0(g^*-h)}{\rho_{0i} h}}{p} (-1)^{N-p-2}  = 0,
    \end{align}
    which for $h=0$ is 
    \begin{equation}
    	 \prod_i \frac{\rho_{0i} g_i - \rho^*_0 g}{\rho_{0i}} = 0,  
    \end{equation} which is not possible since $\rho^*_0 g^*\neq \rho_{0i} g_i$. Consequently when $m=1,2$ the characteristic polynomial cannot be permanently degenerate.

 \section{Proof of Theorem \ref{thm:A:vec}} \label{p-thm:A:vec} 
    From Theorem \ref{thm:A:vectors}, it suffices to find eigenvectors for the matrix $\alpha\rho$ with eigenvalues $\lambda^2 = \rho^*_0(g^*_0-h)$. A straightforward computation gives the components of $\alpha\rho-\lambda^2$
    \begin{equation}
    	 (\alpha\rho-\lambda^2)_{ij} = \left\{ \begin{array}{ll} \rho_{0i} g_i - \rho^*_0 g^* + \rho^*_0 h, & i=j,\\ \rho_{0i} h, & i\neq j. \end{array}\right.
    \end{equation}
    Since the matrix $\mathcal A$ is permanently degenerate, $m+1$ pairs of $(\rho_{0i} g_i,\rho_{0i})$ are equal to $(\rho^*_0 g^*,\rho^*_0)$. From this, it follows that $m+1$ columns of the matrix $(\alpha\rho-\lambda^2)$ are parallel. Indeed all elements of such columns are $\rho^*_0 h$. We identify these columns by $i_k,\ k=1,2,\ldots m+1$. Define  the $i-$th element of the vector $q^{(k)}$ by 
    \begin{equation}
    	(q^{(k)})^i = \left\{ \begin{array}{ll} 1, & i=i_1,\\ -1, & i=i_{k+1},\\ 0, & \mbox{else} \end{array}\right. \mbox{ for } k=1,2,\ldots m. 
    \end{equation}Then $\alpha\rho q^{(k)} = \rho^*_0(g^*-h)q^{(k)}$. Using the construction of the previous theorem, we obtain the associated eigenvector for $\mathcal A$.
	\bibliography{refs}
	\bibliographystyle{apsrev4-1}
\end{document}